%% file: main.tex
\documentclass[lettersize,journal]{IEEEtran}
\usepackage{amsmath,amsfonts}
\usepackage{algorithmic}
\usepackage{algorithm}
\usepackage{array}
\usepackage{textcomp}
\usepackage{stfloats}
\usepackage{url}
\usepackage{verbatim}
\usepackage{graphicx}
\usepackage{cite}
\usepackage{siunitx}
\usepackage{tikz}
\usepackage[utf8]{inputenc}
\usetikzlibrary{shapes.geometric, arrows.meta, positioning, calc}
\PassOptionsToPackage{usenames,dvipsnames}{xcolor}
\usepackage{pgfplots}
\usepackage{pgfplotstable}
\usepackage{booktabs}
\usepackage{adjustbox}
\usepackage[flushleft]{threeparttable}
\usepackage{hyperref}
\usepackage{orcidlink}
\usepackage{multirow}

\usepackage{subcaption}
\usepackage{caption}

\usepackage{flushend}	
\usepgfplotslibrary{groupplots}
\pgfplotsset{compat=1.18}
\usetikzlibrary{calc}

\input{glossar}

\begin{document}

\title{NN-Based Transistor Modeling for Ultra-Fast SPICE Simulation and Standard Cell Library Characterization}

\author{Rodion~Novkin$^{\orcidlink{0009-0006-6632-9804}}$, and Hussam~Amrouch$^{\orcidlink{0000-0002-5649-3102}}$,~\IEEEmembership{Member,~IEEE}

\thanks{
Rodion Novkin and Hussam Amrouch are with the Technical University of Munich; TUM School of Computation, Information and Technology; Chair of AI Processor Design; Munich Institute of Robotics and Machine Intelligence, Munich, Germany (e-mail: \nolinkurl{{rodion.novkin,amrouch}@tum.de}). 
}

}


\markboth{}
{Shell \MakeLowercase{\textit{Novkin et al.}}: NN-Based Transistor Modeling for Ultra-Fast SPICE Simulation and Standard Cell Library Characterization}


\maketitle

\begin{abstract}
\nn-based transistor models have proven to be a promising solution to accelerate device modeling. Although these models demonstrated remarkable speedup in circuit simulations, they were not rigorously tested in complex tasks within standard EDA tool flows for chip design, which require very high transistor model precision to provide accurate results and converge properly. To investigate their full potential and promise, we develop NN-based transistor models that accurately predict the drain current and various charges for a \qty{3}{nm} nanosheet, and employ the models for a standard cell library characterization. This enables us to evaluate the accuracy and speedup under a very wide range of SPICE simulations, benchmarking our models against an industry-standard implementation. Since standard cell library characterization involves thousands of SPICE simulations, it serves as a great benchmark for transistor model quality and consistency. To further evaluate model accuracy, we also perform analog- and digital-level simulations for various complex circuits. Our experiments reveal that NN-based models are a valid alternative to existing transistor compact models, capable of producing accurate delay and power estimates with a much higher speed and convergence time compared to the conventional Verilog-A-based industry-standard (BSIM-CMG) transistor compact model. Our NNs achieve sub-$0.11\%$ and $0.76\%$ errors for delay and total power evaluation on a wide range of circuits, while taking up to 8 times less time during the standard cell library characterization. Additionally, we provide comprehensive information and guidance on NN accuracy, unveiling the relationship between NN size and its effect on circuit simulation precision.

\end{abstract}


\section{Introduction}

\IEEEPARstart{A}{ccurate} transistor models are essential for the semiconductor industry since they are required for a wide variety of complex tasks, such as design and simulation of various circuits. Berkeley short-channel IGFET (BSIM) \cite{Chauhan14, Duarte2015} is the industry-standard model that is commonly used to perform these tasks. BSIM relies on physics-based equations to accurately reproduce underlying transistor behavior, enabling the model to achieve excellent performance in precise modeling of device characteristics. However, developing an analytical solution to accurately describe the underlying physics of device operation is often very time-consuming and requires substantial domain expertise since transistor behavior becomes more and more complex in advanced sub-5nm nodes and emerging technologies, where multiple new effects and challenges attributed to advanced transistor structures start to take place. Furthermore, all new effects must be seamlessly integrated into an existing BSIM code and built upon it without altering the model's prior behavior, which unavoidably introduces additional fitting parameters alongside the hundreds of already existing ones and greatly increases device simulation time. Finally, if the underlying physics were not properly captured, an analytical-based model may not even provide the correct output for the necessary transistor characteristics, leading to faulty circuit behavior.

\nn-based compact transistor modeling received significant attention over the recent years due to \nn's promising ability to learn complex data patterns. NNs can approximate any function \cite{Hornik89} without the need to learn underlying physical processes and can effectively create a completely standalone, compact model for specific working conditions dictated by the range of data and input parameters used for NN training. A simplistic nature of matrix multiplication, which is prevalent in most \nn architectures, provides compact models with the potential to greatly reduce the complexity of the model, thus lowering the required time for circuit simulation, which can be reduced even further with GPU-acceleration of NN-based transistor models and SPICE solvers.

An outstanding performance of a standalone NN-based compact modeling approach was successfully demonstrated in multiple prior works \cite{Tung2022, Tung2023, Wang21, Eom24, Tung24spice, NovkinDualNN25} on a wide range of transistor types and technology. Even though \mlp was a common choice for device characteristics modeling, other architectures, e.g, \kan \cite{Novkin25KAN}, \gnn \cite{Jang23gnn}, and a semantic segmentation network \cite{Thomann2024}, were also explored. In addition to pure NN-based compact modeling, multiple hybrid approaches that utilize NN to correct the output of a low-fidelity equation-based model were also presented \cite{Sheelvardhan24, Kao22, Choe24}. Finally, knowledge-based neural networks (KNBB), which allow NNs to learn quantities through empirical physics-based equations and even enable some post-training calibration, were introduced and expanded upon very recently \cite{Kim20, Li23 ,Tung25penn}.

Most of the NN-based transistor models demonstrated high accuracy and great fitting capabilities for important transistor characteristics like drain and gate currents ($I$), charges ($Q$), and pairing capacitances ($C$), which resulted in accurate circuit simulations of different complexity \cite{Tung2022, Tung2023, Choe24, Tung24spice, NovkinDualNN25, Tung25penn}. However, most circuit simulations using NN-based transistor models demonstrated just a visual match with an industry-standard reference model on a very limited set of circuits. Additionally, most research works were focused exclusively on nMOS while keeping pMOS model source and implementation undisclosed. Therefore, a comprehensive and statistically supported evaluation of NN-based transistor model performance for circuit simulation was never performed. Since standard cell library characterization involves thousands of SPICE simulations across a wide variety of circuits, capacitive loads, and transition slews, it provides a strong benchmark for transistor model robustness, accuracy, and convergence. Standard cell library characterization and \sta represent vital and fundamental SPICE-based tasks in the EDA flow, which require accurate transistor models for precise timing and power evaluation. Machine learning was previously used for aging analysis, standard cell library characterization, and power/delay estimation \cite{Klemme20, Klemme21, Klemme22, Klemme221, Genssler23, Rakesh25, Cao25}. However, NNs were never employed as a model on a transistor level. Thus, it is not clear whether they are capable of producing precise delay and power values for complex circuit simulations and how NN accuracy and size affect the precision of the delay and power estimation later.

\noindent
\textbf{Our main contributions within this paper are as follows:}

        \noindent \textbf{(1)} We demonstrate, for the first time, the speed and accuracy of NN-based transistor models for standard cell library characterization, as well as power and timing signoff estimation, benchmarking nMOS and pMOS NN-based transistor models against industry-standard implementation of the 3nm nanosheet using standard EDA tools and a variety of circuits.
        
        \noindent
        \textbf{(2)} We perform standard cell library characterization employing 71 combinational and sequential cells in 1701 SPICE simulations containing 7$\times$7 slew-capacitive load combinations each, achieving $0.06\%$ and $0.33\%$ average relative errors for delay and cell power estimation among all simulations.
        
        \noindent
        \textbf{(3)} Using the generated libraries, we evaluate timing and power signoff for 12 industrial circuits with the number of gates ranging from 70 to 104286, where NN-based transistor models achieve sub-$0.11\%$ and $0.76\%$ relative errors for delay and total power estimation, respectively.

        \noindent
        \textbf{(4)} We perform a SPICE circuit simulation for a full 16-bit divider and 1001-stage NAND ring oscillator (RO) consisting of 5270 and 4004 transistors, demonstrating a simulation speedup of up to 10x and 43x, respectively.


\section{Methodology for NN-based Transistor Modeling} \label{sec:method}
\begin{figure}
\centering
\includegraphics[height=10cm, keepaspectratio]{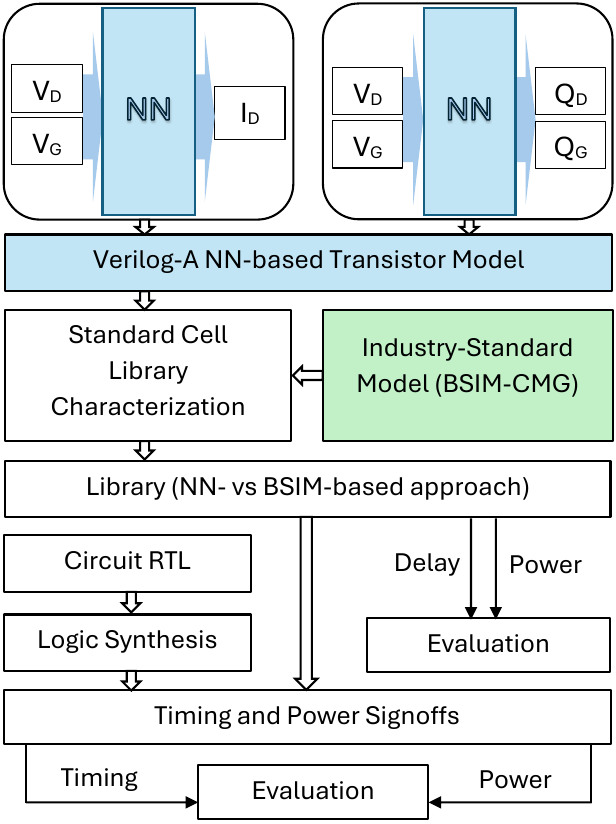}
    \caption{NN-based transistor model is implemented in Verilog-A out of pre-trained networks. Then, all transistor models are employed for the standard cell library characterization, and the output libraries are used for timing and power signoff estimation and compared to the industry-standard BSIM-CMG.} 
\label{fig:flow}
\end{figure}

In the following, we describe the selected device and dataset generation process. NN architecture, used for transistor modeling, and the training algorithm (including conversion, loss, and error functions) are presented in Section \ref{sub:nnta}. Simulation setup for the standard cell library characterization and circuit analysis is detailed in Section \ref{sub:simsetup}. A complete experimental workflow is outlined in Fig. \ref{fig:flow}.

\subsection{Dataset Generation} \label{sub:dataset}
For systematic evaluation of the NN-based transistor models, a \qty{3}{\nano\meter} nanosheet with a regular threshold voltage was selected from an open-source PDK \cite{gt3}. For a single device, we model the drain current \ids, drain charge \qd, and gate charge \qg characteristics by varying the drain voltage \vds and gate voltage \vgs. The bulk charge \qb is not considered since it is extremely small compared to other charges and does not contribute any noticeable pairing capacitances for the selected transistor. The source charge is not modeled and calculated as $Q_S=-(Q_D+Q_G)$ to maintain charge conservation. Training data is generated with a voltage step of \qty{5}{mV} on a range from \qty{-0.2}{V} to \qty{1}{V} for nMOS and from \qty{-1}{V} to \qty{0.2}{V} for pMOS. Opposite voltage regions (negative for nMOS and positive for pMOS) are mainly generated to account for overshoots that are common in circuit simulations. Additionally, to learn the behavior of derivatives correctly with NN, it is necessary to extend the boundaries of the data so that they slightly exceed the intended operating range. In the scope of this work, the intended operating range, including overshoots, is from \qty{-0.1}{V} to \qty{0.9}{V} for nMOS and from \qty{-0.9}{V} to \qty{0.1}{V} for pMOS. Test data was generated with a slight shift in a range from \qty{-0.097}{V} to \qty{0.893}{V} for nMOS and in a range from \qty{-0.893}{V} to \qty{0.097}{V} for pMOS with the same \qty{5}{mV} step. In addition to drain current and charge values, their first and second order derivatives with respect to \vds and \vgs are also calculated since they are required for training. All data was obtained with a commercial SPICE simulator using BSIM-CMG for gate-all-around transistors (version 111.2.1) \cite{bsim111}.

\subsection{Neural Network Architecture and Training Algorithm} \label{sub:nnta}
Since our goal is to explore the relation between NN model size and its precision for circuit simulation, we employ two different MLPs with the same architecture, composed of input, output, and two hidden layers connected with the $Tahn$ activation function. A smaller NN features a hidden dimension of 8 neurons, while a bigger NN has 16. In total, we employ two NNs for each size: the drain current NN output $y_I$ produces a magnitude of the drain current using drain voltage \vds and gate voltage \vgs as inputs; and the output of the charge NN $y_Q$ produces scaled values for \qd and \qg with the same inputs. The final values for \ids and both charges are calculated with the following conversion functions:
\begin{equation}\label{eqn:cf1}
    I_D = V_D \cdot e^{y_I}
\end{equation}
\begin{equation}\label{eqn:cf2}
    Q_{D,G} = y_Q \cdot \lambda
\end{equation}
Similar to other works \cite{Tung2022,Tung2023,Eom24}, the sign of the drain current is dictated by the sign of \vds, while the signs of the charges have to be learned since they cannot be easily determined. The charge scaling factor $\lambda$ in Equation \ref{eqn:cf2} helps to keep NN output values within a reasonable range and is based on the maximum charge value of a transistor.

All networks are trained using the PyTorch framework \cite{pytorch} and weighted loss functions with various coefficients that adjust the influence of different loss function terms:
\begin{equation}\label{eqn:loss1}
\begin{split}
    L_I = a\cdot Er(I_D) + Er(y_I) + b\cdot Er(g_m) + \\ + c\cdot Er(g_{DS}) + d\cdot Er(g'_m) + e\cdot Er(g'_{DS})
\end{split}
\end{equation}
\begin{equation}\label{eqn:loss2}
\begin{split}
    L_Q = a\cdot Er(Q) + b\cdot Er(\frac{\partial Q}{\partial V_G}) + \\ + c\cdot Er(\frac{\partial Q}{\partial V_D}) + d\cdot Er(\frac{\partial^2 Q}{\partial V_D^2}) + e\cdot Er(\frac{\partial^2 Q}{\partial V_G^2})
\end{split}
\end{equation}
\begin{equation}\label{eqn:mae}
    Er(Y)=\frac{1}{p}\sum_{i=1}^p|y_i-\hat{y_i}|
\end{equation}
We employ mean absolute error (MAE) as an error function to avoid error spikes when training NNs with scaled charges and to keep drain current errors reasonably high since they are not scaled. Training is conducted in an iterative manner, meaning that different loss terms are enabled sequentially. 

All NNs are trained with an Adam optimizer \cite{adam} and an initial learning rate of 0.001, which is adjusted with the learning rate scheduler during training. For the drain current \ids, \textbf{1)} the training starts with just fitting the margin error $L_I = Er(y_I)$. \textbf{2)} In the next step, coefficients $a$ and $b$ are enabled to ensure better fitting in the saturation region and to add the transconductance component. \textbf{3)} Then, an output conductance loss term $c$ is introduced. Finally, \textbf{4)} both second-order derivatives are enabled with coefficients $d,e$, and after that \textbf{5)} NNs are additionally trained with a complete loss function and reduced initial learning rate.

Since the precision of capacitances is more relevant than the precision of charges themselves, \textbf{1)} charge NN training starts with the loss $L_Q = Er(\frac{\partial Q}{\partial V_G})$. \textbf{2)} In the next step, another first-order partial derivative is enabled with $c$ followed by \textbf{3)} an introduction of the second-order derivatives with coefficients $d$ and $e$. Then, \textbf{4)} an actual charge component is added with coefficient $a$. Finally, \textbf{5)} NNs are additionally trained with a complete loss function and reduced initial learning rate, similar to drain current networks.

Outside of conversion functions shown in Equation \ref{eqn:cf1} and \ref{eqn:cf2}, neither input nor output data is normalized in any other way to avoid additional computation during gradient calculation, which is necessary for the loss functions. For the same reason, no normalization layers are used in the NN architecture. The values of the loss term coefficients are selected to ensure a reasonably equal contribution from different loss terms. However, this rule is not very strict, and the coefficients are rather flexible when used with the state-of-the-art NN optimizers. Depending on the task's focus, more important terms can be given a stronger influence. However, to remain helpful, the loss contribution of any term should not be hundreds or thousands of times smaller than that from other terms.

\begin{table}
\centering
\caption{Circuits used for timing and power signoff estimation.}
\label{table:sta_circuits}
\resizebox{\columnwidth}{!}{%
\begin{tabular}{c|c|c}
\toprule
Abbreviation &  Description & Cells\\ \midrule
Norm &  Normalization for Fractional Input & 70\\
CRC &  Cyclic Redundancy Check & 150\\
Adder128 & 128-bit Adder & 194 \\
InvSqrtFP &  Floating-Point Reciprocal of Square Root & 1271\\
ExpFP &  Floating-Point Exponential & 4214\\
SinCosFP &  Floating-Point Sine or Cosine & 5017\\
MAC32 &  32-bit Multiply-Accumulate & 6512\\
Mult64 &  64-bit Multiplier & 19898\\
Exp64 &  64-bit Exponential & 65914\\
Div64 &  64-bit Divider & 72304\\
RISC-V &  64-bit RISC-V & 6529\\
MAC-array &  16x16 8-bit MAC-Array for AI acceleration & 104286\\
\bottomrule
\end{tabular}}
\end{table}

\subsection{Simulation Setup for Standard Cell Library Characterization and Circuit Analysis} \label{sub:simsetup}
In our library, we employ 71 combinatorial and sequential standard cells, with each cell characterized at rising and falling edges. The function and behavior of each cell are described by the measurement information about timing, power, leakage, and area. During the standard cell library characterization process, each cell is simulated and analyzed for 7 signal slews and 7 load capacitances, resulting in 49 total combinations. Full standard cell library characterization employs 1701 post-layout SPICE netlists, including parasitics. The library is essentially a database containing all data from these simulations. It has a well-defined format and serves as an input for further circuit analysis, synthesis, and timing/power estimation.

In order to evaluate how model accuracy scales from the individual cell level to the circuit level, we perform \sta for 12 different circuits, described in Table \ref{table:sta_circuits}. To cover a wide range of applications and data types, we selected 10 industrial arithmetic circuits, RISC-V (Rocket) \cite{chipyard}, and an in-house AI-accelerator, which consists of 16x16 systolic MAC-arrays. All circuits were synthesized with a supply voltage $V_{dd}$ = \qty{0.8}{V}, resulting in a range of 70 to 104286 standard cells.

\section{Accuracy Evaluation at the Transistor Level} \label{sec:acc_tr}
Before conducting any experiments and performing measurements, all trained neural networks are coded into a standalone transistor model written in Verilog-A. Similar to other works (e.g. \cite{Tung24spice}), we unroll all the loops within matrix multiplications. All data, shown in this section, is obtained from commercial EDA tools using different transistor models: industry-standard BSIM-CMG and our two NN-based transistor models (NN16, NN8), where the number indicates the number of neurons in the hidden layers.
\begin{figure*}
\centering
\begin{subfigure}[b]{0.9\textwidth} 
    \includegraphics[height=4cm, keepaspectratio]{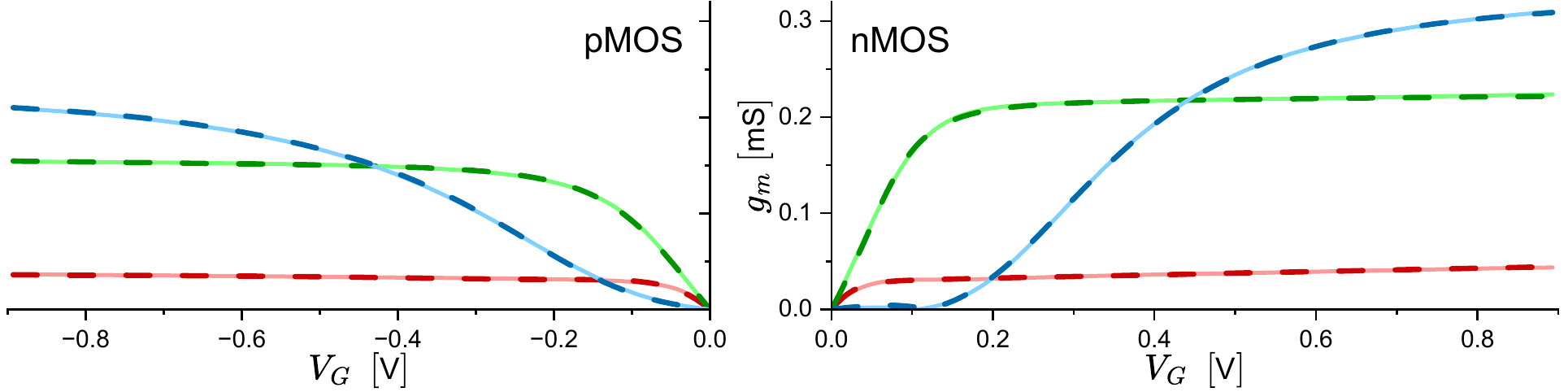}
    \label{fig:iv_plots1}
\end{subfigure}\hspace{0.01em}%
\begin{subfigure}[b]{0.9\textwidth} 
    \includegraphics[height=3.8cm, keepaspectratio]{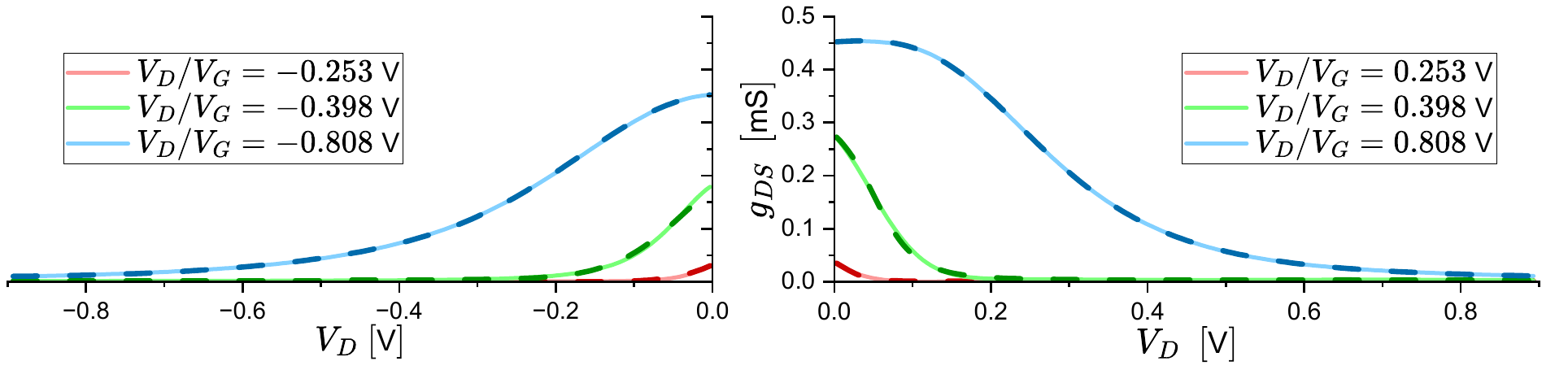}
    \label{fig:iv_plots2}
\end{subfigure}\hspace{0.01em}%

\medskip
    \caption{NN8 model exhibits an excellent fit for nMOS and pMOS transconductance and output conductance. Lines are NN8 outputs, similar-colored dashed lines - BSIM-CMG industry-standard compact model. Either \vgs or \vds is fixed, depending on the plot. Y-scale is the same for nMOS (right) and pMOS (left) plots.} 
\label{fig:iv_plots}
\end{figure*}

\begin{figure}
\centering
\begin{subfigure}[b]{0.44\textwidth} 
    \includegraphics[height=2.7cm, keepaspectratio]{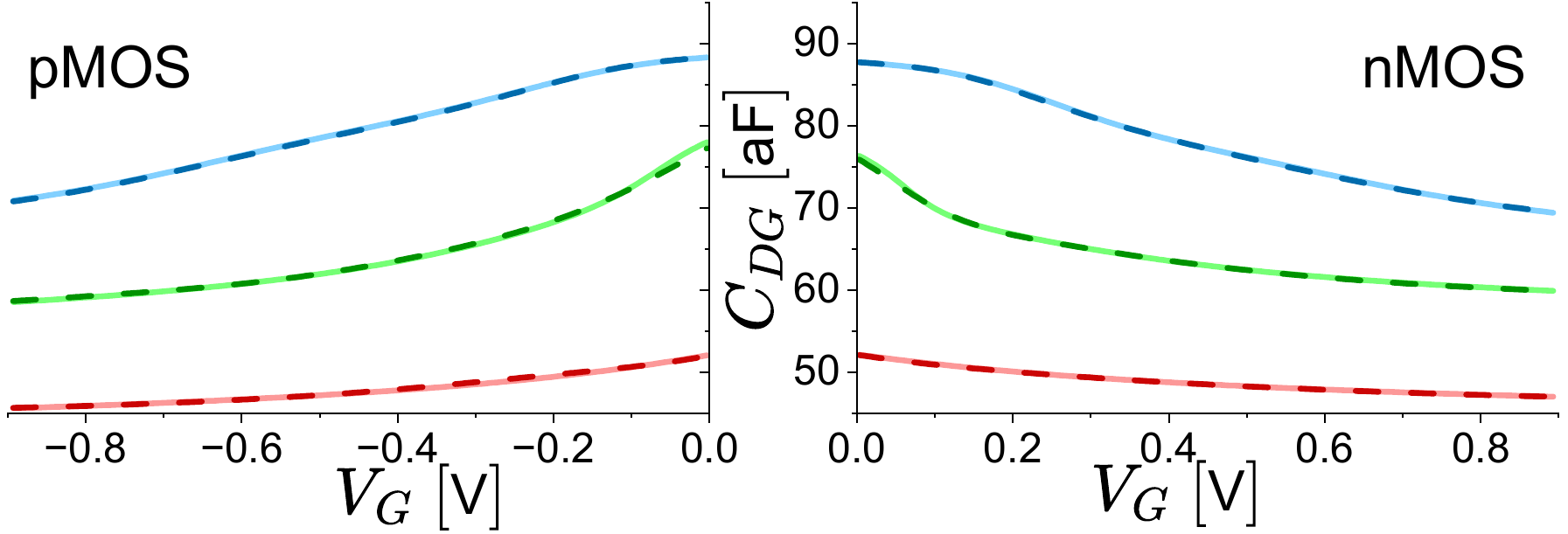}
    \label{fig:cv_plots1}
\end{subfigure}\hspace{0.01em}%
\begin{subfigure}[b]{0.44\textwidth} 
    \includegraphics[height=2.74cm, keepaspectratio]{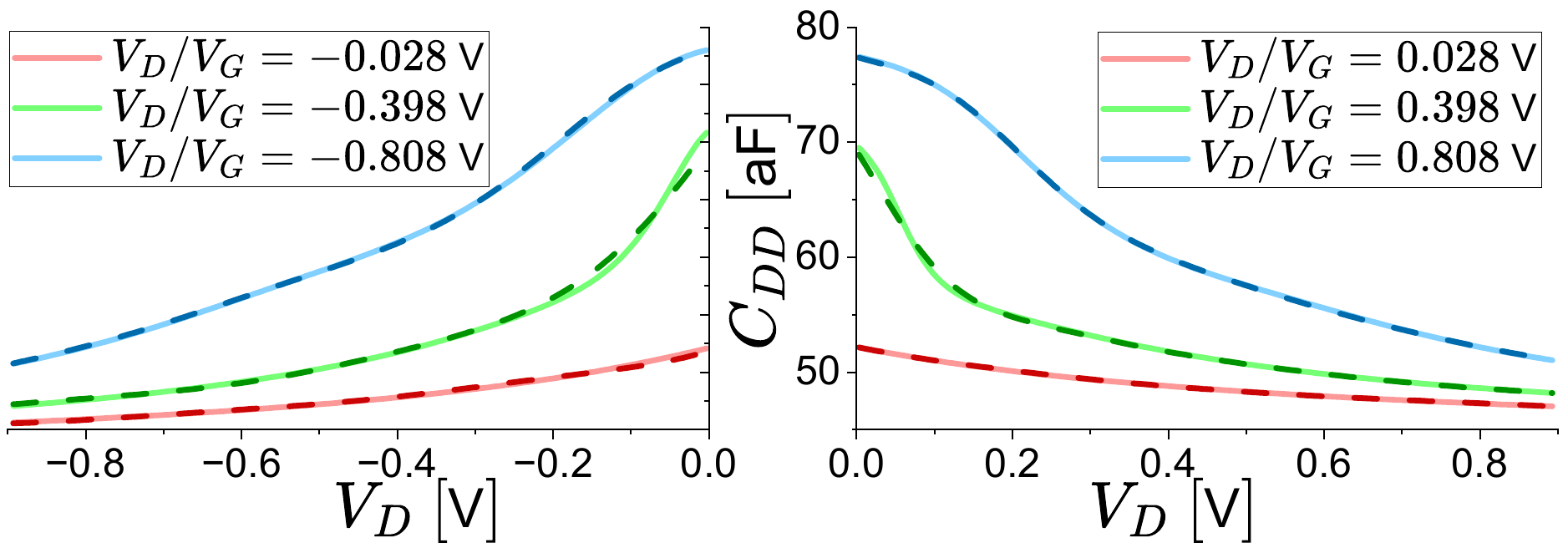}
    \label{fig:cv_plots2}
\end{subfigure}\hspace{0.01em}%
\begin{subfigure}[b]{0.43\textwidth} 
    \includegraphics[height=2.72cm, keepaspectratio]{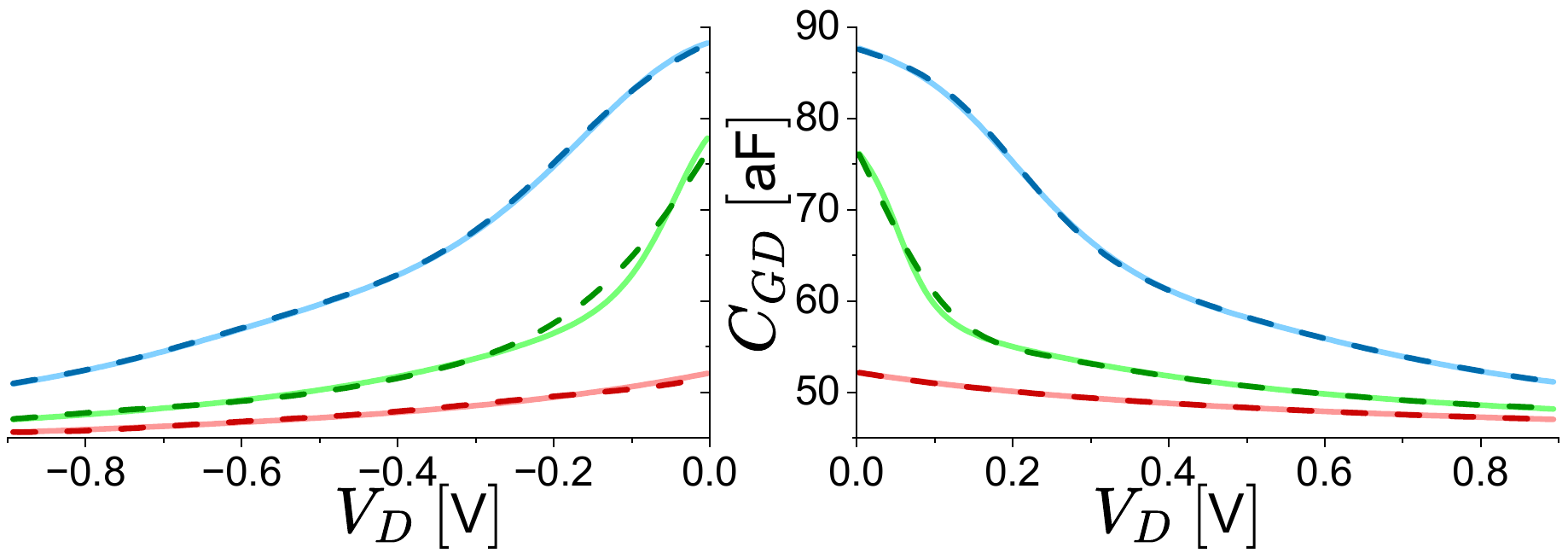}
    \label{fig:cv_plots3}
\end{subfigure}\hspace{0.01em}%
\begin{subfigure}[b]{0.44\textwidth} 
    \includegraphics[height=2.72cm, keepaspectratio]{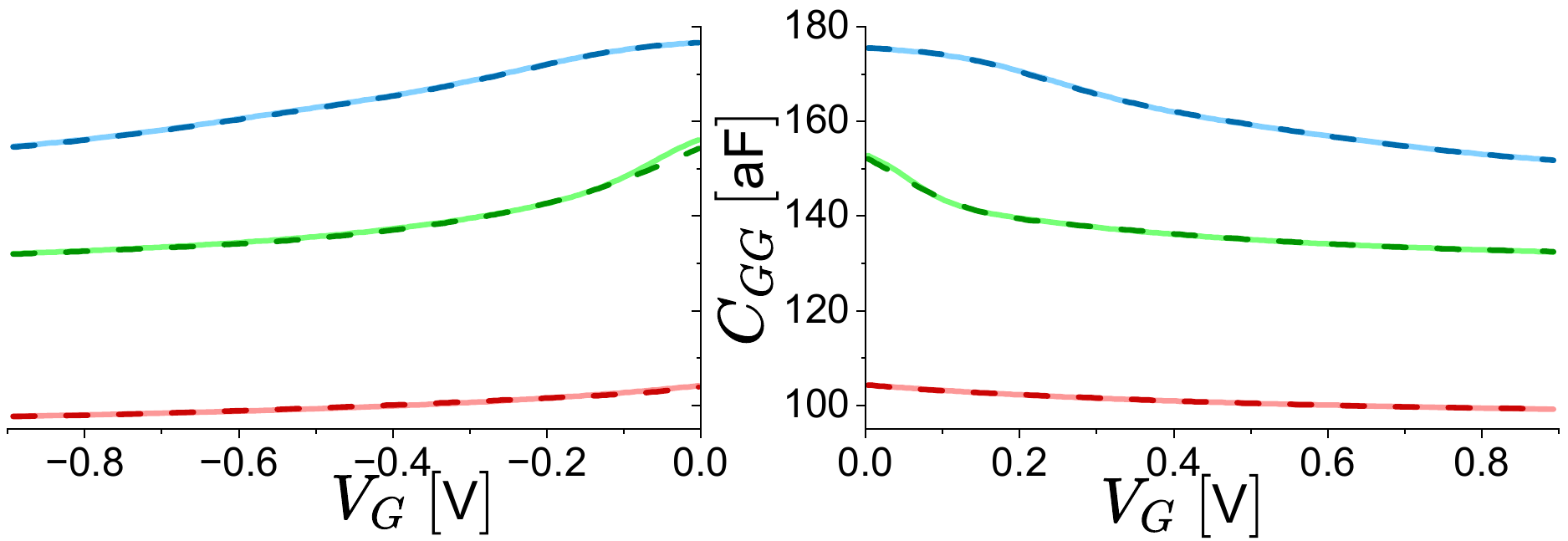}
    \label{fig:cv_plots4}
\end{subfigure}

\medskip
    \caption{NN8 model demonstrates a good fit for capacitances in necessary regions. Lines are NN8 outputs, similar-colored dashed lines - BSIM-CMG industry-standard compact model. Either \vgs or \vds is fixed, depending on the plot. Y-scale is the same for nMOS (right) and pMOS (left) plots.} 
\label{fig:cv_plots}
\end{figure}

We perform a comprehensive evaluation of the accuracy for all NN-based transistor models using a wide range of characteristics necessary for precise circuit simulations. All visual plots for characteristics are shown for both nMOS and pMOS on the same y-scale to highlight their asymmetry. To avoid clutter, NN16 was omitted from the visual evaluation, as it produces even more precise plots than NN8. Fig. \ref{fig:iv_plots} demonstrates an excellent fit for transconductance \gm and output conductance \gds produced by the 8-neuron NN-based model. The conductance behavior is captured very well throughout the entire testing voltage range. Similar quality can be observed in all capacitance plots shown in Fig. \ref{fig:cv_plots}, where NN8 is able to capture the behavior of various capacitances with great accuracy in the most important operating regions, i.e., near-zero voltage and near the intended supply voltage equal to \qty{0.8}{V}. However, in the region near \qty{0.4}{V}, NN8 struggles to capture charge derivative behavior properly, and this observation will be important later. 

Although visual estimation is extremely beneficial for a general evaluation of NN-based transistor model performance, it does not provide a full picture of how accurate every NN is. Thus, we employ two quantitative metrics to evaluate absolute and relative accuracy of all important characteristics - \mape and \rmse, calculated as:
\begin{equation}\label{eqn:mape}
    \text{\mape}=\frac{\sum_{i=1}^{n} |y_i-\hat{y}_i|}{\sum_{i=1}^{n}|y_i|}
\end{equation}

\begin{equation}\label{eqn:rmse}
    \text{\rmse}=\sqrt{\frac{\sum_{i=1}^{n} (y_i-\hat{y}_i)^2}{n}}
\end{equation}

Table \ref{table:accuracy} provides an overview of the NN-based model test accuracy compared to the BSIM-CMG, which was used to generate train and test data. All capacitances not listed in the table are several magnitudes lower than the listed ones and do not significantly affect circuit simulations. As expected, bigger NNs demonstrate better results than their smaller counterparts for all important characteristics. Since the pMOS data pattern is different from the nMOS, same-size NNs did not achieve the same accuracy for different transistor types. While \rmse values are easily interpretable, \mape ones have more depth and may require additional context. The majority of data are on an NN-friendly scale, meaning the ground truth and predicted values are neither extremely low or high on a linear scale. In contrast, \gm and \gds values can get really close to zero, resulting in rather high average relative errors (e.g., a $100\%$ relative error from $\hat{y}_i=0.002$ and $y_i=0.001$ significantly contributes to the average value, especially when the average is very low). Thus, \mape for these characteristics is noticeably higher than in other cases, which skews the perception of the results but barely contributes to an actual performance of the NN-based models, since \rmse for these cases is still low.

\begin{table*}
\centering
\caption{Prediction error of trained neural networks.}
\label{table:accuracy}
\resizebox{0.85\textwidth}{!}{%
\begin{tabular}{c|cc|cc|cc|cc}
\toprule
\multirow{3}{*}{Variable} & \multicolumn{4}{c|}{nMOS} & \multicolumn{4}{c}{pMOS}\\
& \multicolumn{2}{c}{NN16} & \multicolumn{2}{c|}{NN8} & \multicolumn{2}{c}{NN16} & \multicolumn{2}{c}{NN8}\\
& RMSE & MAPE & RMSE & MAPE & RMSE & MAPE & RMSE & MAPE\\ \midrule
\ids     & 0.026 \unit{\micro\ampere} & 0.32 $\%$ & 0.056 \unit{\micro\ampere} & 0.73 $\%$ & 0.037 \unit{\micro\ampere}  & 0.51 $\%$ & 0.072 \unit{\micro\ampere} & 0.59 $\%$ \\ 
$g_m$    & 0.290 \unit{\micro\siemens} & 2.54 $\%$ & 0.914 \unit{\micro\siemens} & 8.84 $\%$ & 0.369 \unit{\micro\siemens}  & 0.89 $\%$ & 0.553 \unit{\micro\siemens} & 1.03 $\%$ \\ 
$g_{DS}$ & 0.677 \unit{\micro\siemens} & 4.28 $\%$ & 1.256 \unit{\micro\siemens} & 8.54 $\%$ & 0.737 \unit{\micro\siemens}  & 4.42 $\%$ & 1.077 \unit{\micro\siemens} & 7.54 $\%$ \\ 
\qd      & 0.007 \unit{aC}  & 0.22 $\%$ & 0.011 \unit{aC} & 0.29 $\%$ & 0.041 \unit{aC}  & 3.84 $\%$ & 0.036 \unit{aC} & 1.93 $\%$ \\ 
\qg      & 0.012 \unit{aC}  & 0.24 $\%$ & 0.021 \unit{aC} & 0.24 $\%$ & 0.091 \unit{aC}  & 0.69 $\%$ & 0.070 \unit{aC} & 0.42 $\%$ \\ 
\qs      & 0.008 \unit{aC}  & 0.09 $\%$ & 0.014 \unit{aC} & 0.12 $\%$ & 0.051 \unit{aC}  & 0.19 $\%$ & 0.043 \unit{aC} & 0.53 $\%$ \\ 
$C_{DG}$ & 0.052 \unit{aF}  & 0.04 $\%$ & 0.111 \unit{aF} & 0.10 $\%$ & 0.102 \unit{aF}  & 0.13 $\%$ & 0.297 \unit{aF} & 0.31 $\%$ \\ 
$C_{DD}$ & 0.098 \unit{aF}  & 0.06 $\%$ & 0.170 \unit{aF} & 0.12 $\%$ & 0.080 \unit{aF}  & 0.08 $\%$ & 0.312 \unit{aF} & 0.27 $\%$ \\ 
$C_{GG}$ & 0.089 \unit{aF}  & 0.03 $\%$ & 0.177 \unit{aF} & 0.07 $\%$ & 0.214 \unit{aF}  & 0.13 $\%$ & 0.571 \unit{aF} & 0.27 $\%$ \\ 
$C_{GD}$ & 0.137 \unit{aF}  & 0.08 $\%$ & 0.211 \unit{aF} & 0.15 $\%$ & 0.132 \unit{aF}  & 0.12 $\%$ & 0.523 \unit{aF} & 0.43 $\%$ \\ 
$C_{GS}$ & 0.138 \unit{aF}  & 0.09 $\%$ & 0.224 \unit{aF} & 0.17 $\%$ & 0.256 \unit{aF}  & 0.27 $\%$ & 0.606 \unit{aF} & 0.52 $\%$ \\ 
$C_{SG}$ & 0.050 \unit{aF}  & 0.04 $\%$ & 0.106 \unit{aF} & 0.10 $\%$ & 0.122 \unit{aF}  & 0.13 $\%$ & 0.373 \unit{aF} & 0.39 $\%$ \\ 
$C_{SS}$ & 0.093 \unit{aF}  & 0.07 $\%$ & 0.181 \unit{aF} & 0.16 $\%$ & 0.141 \unit{aF}  & 0.16 $\%$ & 0.439 \unit{aF} & 0.45 $\%$ \\ \bottomrule
\end{tabular}}
\end{table*}

Going from 16 neurons to 8, nMOS NNs lose roughly half of the absolute precision. pMOS NNs experience a similar accuracy drop for the drain current, but capacitance absolute errors increase by almost three times. The comparison between nMOS and pMOS NNs shows that pMOS NN16 achieves a similar accuracy as nMOS NN8 which indicates that the data pattern for pMOS capacitances is harder to learn. Therefore, for the same NN capacity (i.e., the number of weights), the fitting capabilities will differ for different devices. When NN capacity is further reduced by lowering the number of neurons in hidden layers, the pMOS charge model is more impacted and experiences a greater drop in capacitance accuracy, as it was already struggling to fit the data on the same level as nMOS. In contrast to a regular function regression, NNs in transistor models not only have to learn the function output, but its derivatives with respect to multiple variables as well. Therefore, if NN does not have a sufficient number of parameters to learn everything, the optimizer can select an easier-to-learn loss term to optimize during the training process, which may result in a situation demonstrated by pMOS charge NN - for NN8 charge accuracy goes slightly up compared to NN16, but capacitance accuracy decreases significantly. Although it is possible to balance loss terms with coefficients to mitigate this effect and potentially enhance capacitance accuracy, we will keep the charge NN8 for pMOS unchanged to assess how larger errors impact circuit simulation and further analysis.

\section{Analog Level Circuit Simulation} \label{sec:circuit}
In this section, we perform analog level analysis using SPICE simulations for two circuits, comparing precision and speedup of NN-based models with respect to BSIM-CMG.
\subsection{Accuracy of Analog Circuits}
We perform a comprehensive evaluation of the transistor model performance for a 16-bit divider circuit designed with 5270 nMOS and pMOS transistors in total. The simulation is conducted with the same netlist and various transistor models in SPICE using two setups representing the precision of the circuit simulation. The default simulation setup involves low tolerances, as well as a predefined optimization profile that trades off the simulation speed for a much higher accuracy. This setup is commonly used when a very precise simulation is required for an accurate delay estimation. The second simulation setup is a speed-oriented one, featuring higher tolerances and a predefined optimization profile that trades off simulation accuracy for higher speed and faster convergence. The latter setup will be referred to as a low-precision simulation (LPS) in all further experiments. LPS is primarily used to demonstrate its immense benefit for NN-based modeling speedup, as well as unavoidable limitations for precise circuit simulation, which come with the low precision.

\begin{figure*}
\centering
\includegraphics[height=3.5cm, keepaspectratio]{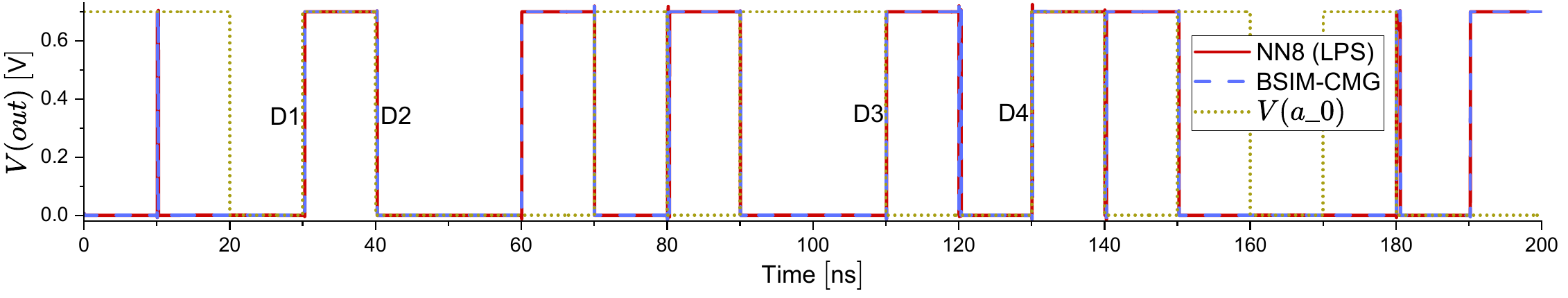}
    \caption{16-bit divider circuit containing 5270 transistors. Simulation with NN8 model and a low-precision simulation (LPS) setup shows an excellent visual match with BSIM-CMG simulated with the default precision setup. D1-D4 indicate the points for delay measurement for the output voltage at \textit{quotient\_0} node with respect to the input \textit{a\_0}. $V_{dd}$ = \qty{0.7}{V}.} 
\label{fig:div16}
\end{figure*}

\begin{figure*}
\centering
\includegraphics[height=2.28cm, keepaspectratio]{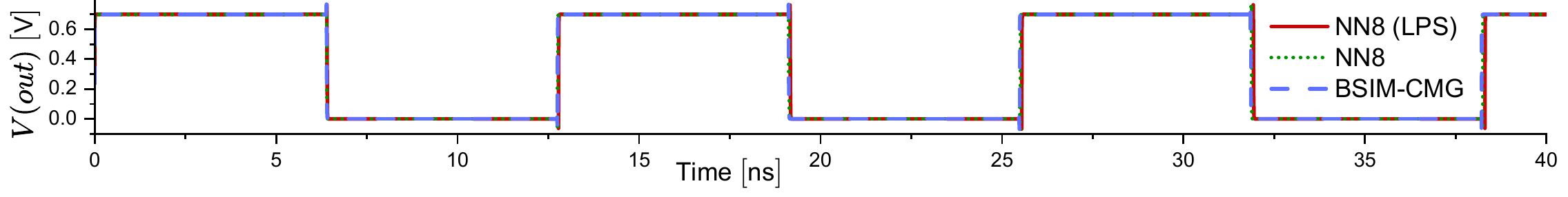}
    \caption{1001-stage NAND ring oscillator (RO) circuit containing 4004 transistors. Simulation with NN8 model and a low-precision simulation (LPS) setup shows minor visual deviations with BSIM-CMG and NN8 simulated with the default precision setup. $V_{dd}$ = \qty{0.7}{V}.} 
\label{fig:osc}
\end{figure*}

\begin{table*}
\centering
\caption{Delay measurements for the 16-bit divider for delay points shown in Fig. \ref{fig:div16} and 1001-stage NAND ring oscillator (RO) period measurements. Relative error (RE) is calculated with respect to BSIM-CMG.}
\label{table:div}
\resizebox{0.85\textwidth}{!}{%
\begin{tabular}{c|c|cc|cc|cc}
\toprule
{Divider} & BSIM-CMG & \multicolumn{2}{c|}{NN16} & \multicolumn{2}{c|}{NN8} & \multicolumn{2}{c}{NN8 (LPS)}\\
{delay point} & Delay [\unit{ps}] & Delay [\unit{ps}] & RE [\%] & Delay [\unit{ps}] & RE [\%] & Delay [\unit{ps}] & RE [\%] \\ \midrule
D1 & \textcolor{black}{296.00} & \textcolor{black}{295.93} & \textcolor{black}{-0.024} & \textcolor{black}{295.66} & \textcolor{black}{-0.115} & \textcolor{black}{295.37} & \textcolor{black}{-0.213} \\
D2 & \textcolor{black}{216.00} & \textcolor{black}{216.07} & \textcolor{black}{+0.032} & \textcolor{black}{216.06} & \textcolor{black}{+0.028} & \textcolor{black}{215.93} & \textcolor{black}{-0.032} \\
D3 & \textcolor{black}{91.40} & \textcolor{black}{91.41} & \textcolor{black}{+0.011} & \textcolor{black}{91.29} & \textcolor{black}{-0.120} & \textcolor{black}{91.19} & \textcolor{black}{-0.230} \\
D4 & \textcolor{black}{51.10} & \textcolor{black}{51.07} & \textcolor{black}{-0.059} & \textcolor{black}{50.99} & \textcolor{black}{-0.215} & \textcolor{black}{50.77} & \textcolor{black}{-0.646} \\ \midrule
RO period [ps]& \textcolor{black}{12720} & \textcolor{black}{12732} & \textcolor{black}{+0.094} & \textcolor{black}{12734} & \textcolor{black}{+0.110} & \textcolor{black}{12752} & \textcolor{black}{+0.252} \\
\bottomrule
\end{tabular}}
\end{table*}

Fig. \ref{fig:div16} demonstrates that even with the smallest and the least accurate NN8 model, combined with the speed-oriented simulation setup, the output of the circuit shows a perfect visual match with the default high precision simulation using an industry-standard BSIM-CMG model. NN-based model accurately represents all fronts and overshoots for an entire \qty{200}{ns} time-frame and can provide all necessary visual information on circuit behavior. However, accurate delay estimation cannot be performed on the shown in Fig. \ref{fig:div16} output since the difference is way beyond the time scale of the plot. Thus, we measure delay values in several points at the $V_{dd}/2$ level, indicated by $D1-D4$ in Fig. \ref{fig:div16}. Table \ref{table:div} demonstrates that the delay produced by NN16 is extremely close to the BSIM-CMG model and exhibits only some negligible deviation due to numerical differences and minor fitting errors. NN8 demonstrates a slightly higher error on average, and the output becomes less stable with relative errors ranging from $0.028\%$ to $0.215\%$ for the divider circuit. Finally, NN8 (LPS) exhibits significantly higher errors compared to other setups, despite providing an excellent visual match for the circuit output. However, based on the comparison with NN8, the low-precision setup itself is a major source of these errors.

The impact of the LPS becomes more noticeable in the 1001-stage NAND ring oscillator (RO) - a circuit with 4004 total number of transistors, which essentially multiplies an error from a single NAND cell. Fig. \ref{fig:osc} demonstrates that NN8 (LPS) starts to deviate slightly from the reference BSIM-CMG output after a certain period of time, while NN8 with the default simulation setup is more precise. Table \ref{table:div} shows a much higher oscillation period difference for NN8 (LPS) compared to the default setup, and the visual deviation is primarily driven by this accumulating over time difference. Nevertheless, all NN-based transistor models eventually deviate from the reference output for the 1001-stage NAND RO, as even the error caused by small numerical differences is multiplied and accumulated over time. Results of the analog level circuit simulations demonstrate that even very small NN-based transistor models can provide high-quality visual output for circuit behavior. However, the impact of NN size and simulation precision on accurate delay estimation is noticeable and needs to be considered.

\subsection{Circuit Simulation Speedup of NN-based Models} \label{sec:circuit_speed}
Another key aspect of the precision-speed balance is the simulation speed. Table \ref{table:analog_time} shows the time that was spent on the simulation of the 16-bit divider circuit with various transistor models and simulation setups. Verilog-A implementation of BSIM-CMG, selected for a fair comparison with Verilog-A NN implementation, has the highest simulation time of \qty{1573}{s} due to having thousands of lines of code and hundreds of complex functions. NNs need only \qty{657}{s} and \qty{272}{s} for 16- and 8-neuron models, respectively. Notably, the number of iterations for both NN simulations is very similar, while Verilog-A BSIM-CMG did roughly 4000 fewer iterations, resulting in an extremely low computational throughput compared to NN models. Computational throughput represents the number of total iterations performed in a fixed amount of time. This relation changes even more when a low-precision simulation setup is used, and NN speed becomes more than $25\times$ and $10\times$ higher than Verilog-A BSIM-CMG for NN8 and NN16, respectively. Further analysis reveals that NN8 (LPS) was able to perform approximately $40\%$ more iterations over the same time, resulting in this extremely high computational speed. However, the number of required iterations is selected by an optimizer and is based on many factors, including model stability, consistency, and derivative smoothness, which smaller NNs can struggle with. Since NN16 is more stable than NN8, it performs similarly to BSIM-CMG in terms of the number of operations in the low-precision case but takes more time due to its bigger size.

In contrast to the 16-bit divider, 1001-stage NAND RO demonstrates even larger speedups compared to BSIM-CMG. Results in Table \ref{table:analog_time} show that computational throughput for NN16 and NN8 is almost $40\times$ and $43\times$ higher than BSIM-CMG, respectively. When a low-precision simulation setup is introduced, the computational throughput ratio remains mostly the same, with a slight increase for both NNs.

Simulation time reduction in Fig. \ref{fig:time_reduction} follows similar pattern as computational throughput. Low-precision simulation setup significantly reduces simulation time with NNs for the 16-bit divider, while 1001-stage NAND RO introduces only a minor change in relative time reduction. Based on consistent results across multiple experiments, it is evident that NN16 achieves similar accuracy to BSIM-CMG while maintaining superior computational efficiency in Verilog-A.

\begin{table*}
\centering
\caption{Time and number of iterations required for the 16-bit divider and 1001-stage NAND ring oscillator (RO) simulations.}
\label{table:analog_time}
\resizebox{\textwidth}{!}{%
\begin{tabular}{c|c|cc|c|c|cc|c}
\toprule
& \multicolumn{4}{c|}{16-bit divider} & \multicolumn{4}{c}{1001-stage NAND RO}\\ \midrule
\multirow{2}{*}{Model} & Time & \multicolumn{2}{c|}{Iterations} & Computational & Time & \multicolumn{2}{c|}{Iterations} & Computational\\ 
 & [s] & Total & Converged & throughput [iter/s] & [s] & Total & Converged & throughput [iter/s]\\ \midrule
BSIM-CMG & 1573 & 29609 & 16198 & \textcolor{black}{18.82} & 5165 & 139191 & 70313 & \textcolor{black}{27}\\
NN16 & 657 & 33635 & 20087 & \textcolor{black}{51.19} & 146 & 156789 & 88929 & \textcolor{black}{1074}\\
NN8 & 272 & 33518 & 20016 & \textcolor{black}{123.23} & 130 & 149391 & 86996 & \textcolor{black}{1149}\\ \midrule
BSIM-CMG (LPS) & 794 & 20094 & 6383 & \textcolor{black}{25.31} & 2615 & 89584 & 28380 & \textcolor{black}{34.3}\\
NN16 (LPS) & 76 & 20619 & 6399 & \textcolor{black}{271.30} & 61 & 84830 & 26386 & \textcolor{black}{1390}\\
NN8 (LPS) & 44 & 28667 & 6345 & \textcolor{black}{651.52} & 53 & 85563 & 26386 & \textcolor{black}{1614}\\
\bottomrule
\end{tabular}}
\end{table*}



\begin{figure}[htbp]
    \centering
    \begin{subfigure}[b]{0.48\textwidth}
        \centering
        \caption{Default precision}
        \begin{tikzpicture}
            \begin{axis}[
                ybar,
                width=\textwidth, 
                height=3.5cm,
                bar width=30pt, 
                ylabel={Speedup [$\times$]},
                symbolic x coords={16-bit Divider, RO},
                xtick=data,
                ymin=0, ymax=50,
                enlarge x limits=0.6,
                legend style={at={(0.2,0.65)}, anchor=south, legend columns=-1},
                nodes near coords,
                every node near coord/.append style={font=\small},
                ymajorgrids=true,
                grid style=dashed,
            ]
                \addplot[fill=blue!30, draw=black] coordinates {(16-bit Divider, 2.39) (RO, 35.38)};
                \addplot[fill=red!30, draw=black] coordinates {(16-bit Divider, 5.78) (RO, 39.73)};
                \legend{NN16, NN8}
            \end{axis}
        \end{tikzpicture}
        \label{fig:reduction_a}
    \end{subfigure}
    \hfill 
    \begin{subfigure}[b]{0.48\textwidth}
        \centering
        \caption{Low precision}
        \begin{tikzpicture}
            \begin{axis}[
                ybar,
                width=\textwidth,
                height=3.5cm,
                bar width=30pt,
                ylabel={Speedup [$\times$]},
                symbolic x coords={16-bit Divider, RO},
                xtick=data,
                ymin=0, ymax=62,
                enlarge x limits=0.6,
                legend style={at={(0.2,0.65)}, anchor=south, legend columns=-1},
                nodes near coords,
                every node near coord/.append style={font=\small},
                ymajorgrids=true,
                grid style=dashed,
            ]
                \addplot[fill=blue!30, draw=black] coordinates {(16-bit Divider, 10.45) (RO, 42.87)};
                \addplot[fill=red!30, draw=black] coordinates {(16-bit Divider, 18.05) (RO, 49.34)};
                \legend{NN16, NN8}
            \end{axis}
        \end{tikzpicture}
        \label{fig:reduction_b}
    \end{subfigure}

    \caption{Relative simulation time reduction for NN-based transistor models compared to BSIM-CMG. All relative values are calculated using data from Table \ref{table:analog_time}.}
    \label{fig:time_reduction}
\end{figure}
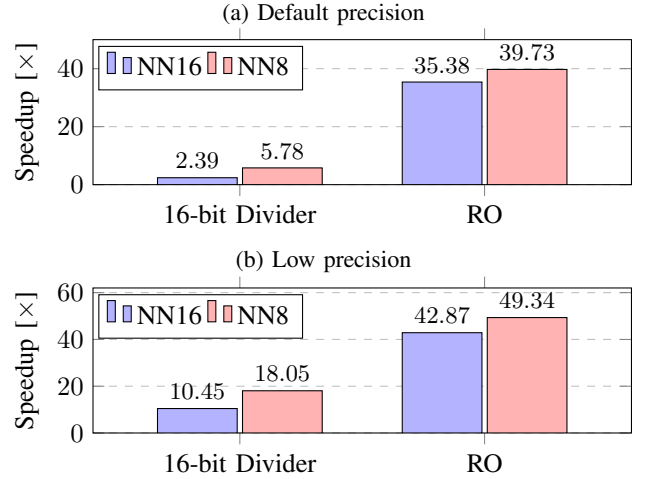

\begin{figure*}
\centering
\begin{tikzpicture}
  \begin{groupplot}[
    group style={group size=4 by 3, horizontal sep=1.0cm, vertical sep=1.3cm},
    width=5.0cm, height=3.0cm,
    grid=major,
    legend style={at={(0.59,-4.35)}, anchor=north, legend columns=-1}
  ]
    \nextgroupplot[title={CRC}]
      \addplot[red, mark=x] coordinates {(0.5,0.357) (0.6,-0.017) (0.7,0.025) (0.8,0.06)};
      \addlegendentry{NN8}
      \addplot[blue, mark=square] coordinates {(0.5,0.19) (0.6,0.059) (0.7,0.076) (0.8,0.022)};
      \addlegendentry{NN16}
    \nextgroupplot[title={ExpFP}]
      \addplot[red, mark=x] coordinates {(0.5,0.307) (0.6,0.057) (0.7,0.047) (0.8,0.09)};
      \addplot[blue, mark=square] coordinates {(0.5,0.183) (0.6,0.084) (0.7,0.082) (0.8,0.066)};
    \nextgroupplot[title={MAC32}]
      \addplot[red, mark=x] coordinates {(0.5,0.275) (0.6,0.077) (0.7,0.13) (0.8,0.2)};
      \addplot[blue, mark=square] coordinates {(0.5,0.171) (0.6,0.095) (0.7,0.09) (0.8,0.094)};
    \nextgroupplot[title={Div64}]
      \addplot[red, mark=x] coordinates {(0.5,0.267) (0.6,0.055) (0.7,0.005) (0.8,0.007)};
      \addplot[blue, mark=square] coordinates {(0.5,0.253) (0.6,0.111) (0.7,0.085) (0.8,0.055)};
    \nextgroupplot[title={Exp64}]
      \addplot[red, mark=x] coordinates {(0.5,0.313) (0.6,0.071) (0.7,0.035) (0.8,0.021)};
      \addplot[blue, mark=square] coordinates {(0.5,0.257) (0.6,0.13) (0.7,0.099) (0.8,0.072)};
    \nextgroupplot[title={Norm}]
      \addplot[red, mark=x] coordinates {(0.5,0.305) (0.6,0.059) (0.7,0.046) (0.8,0.055)};
      \addplot[blue, mark=square] coordinates {(0.5,0.22) (0.6,0.106) (0.7,0.115) (0.8,0.065)};
    \nextgroupplot[title={Adder128}]
      \addplot[red, mark=x] coordinates {(0.5,0.714) (0.6,0.46) (0.7,0.462) (0.8,0.192)};
      \addplot[blue, mark=square] coordinates {(0.5,0.327) (0.6,0.177) (0.7,0.12) (0.8,0.083)};
    \nextgroupplot[title={SinCosFP}]
      \addplot[red, mark=x] coordinates {(0.5,0.227) (0.6,0.02) (0.7,0.078) (0.8,0.13)};
      \addplot[blue, mark=square] coordinates {(0.5,0.141) (0.6,0.069) (0.7,0.07) (0.8,0.058)};
    \nextgroupplot[title={InvSqrtFP}]
      \addplot[red, mark=x] coordinates {(0.5,0.336) (0.6,0.093) (0.7,0.089) (0.8,0.182)};
      \addplot[blue, mark=square] coordinates {(0.5,0.176) (0.6,0.084) (0.7,0.074) (0.8,0.078)};
    \nextgroupplot[title={Mult64}]
      \addplot[red, mark=x] coordinates {(0.5,0.245) (0.6,0.095) (0.7,0.093) (0.8,0.155)};
      \addplot[blue, mark=square] coordinates {(0.5,0.151) (0.6,0.103) (0.7,0.085) (0.8,0.072)};
    \nextgroupplot[title={RISC-V}]
      \addplot[red, mark=x] coordinates {(0.5,0.003) (0.6,-0.146) (0.7,0.04) (0.8,-0.091)};
      \addplot[blue, mark=square] coordinates {(0.5,0.033) (0.6,0.022) (0.7,0.058) (0.8,-0.112)};
    \nextgroupplot[title={MAC-array}]
      \addplot[red, mark=x] coordinates {(0.5,-0.007) (0.6,-0.02) (0.7,0.096) (0.8,0.206)};
      \addplot[blue, mark=square] coordinates {(0.5,0.017) (0.6,0.035) (0.7,0.087) (0.8,0.067)};
  \end{groupplot}

  \draw[thick, -stealth] ($(group c1r3.south west) + (-0.7cm, 0)$) -- 
                         ($(group c1r1.north west) + (-0.7cm, 0)$) 
                         node[midway, above, rotate=90] {Relative error $[\%]$};

  \draw[thick, -stealth] ($(group c1r3.south west) + (0, -0.6cm)$) -- 
                         ($(group c4r3.south east) + (0, -0.6cm)$) 
                         node[midway, below] {Supply voltage $V_{dd}$ $[$\unit{V}$]$};
\end{tikzpicture}

    \caption{Relative error for circuit delay estimation calculated with respect to BSIM-CMG for different supply voltages.} 
\label{fig:sta_over_vdd}
\end{figure*}
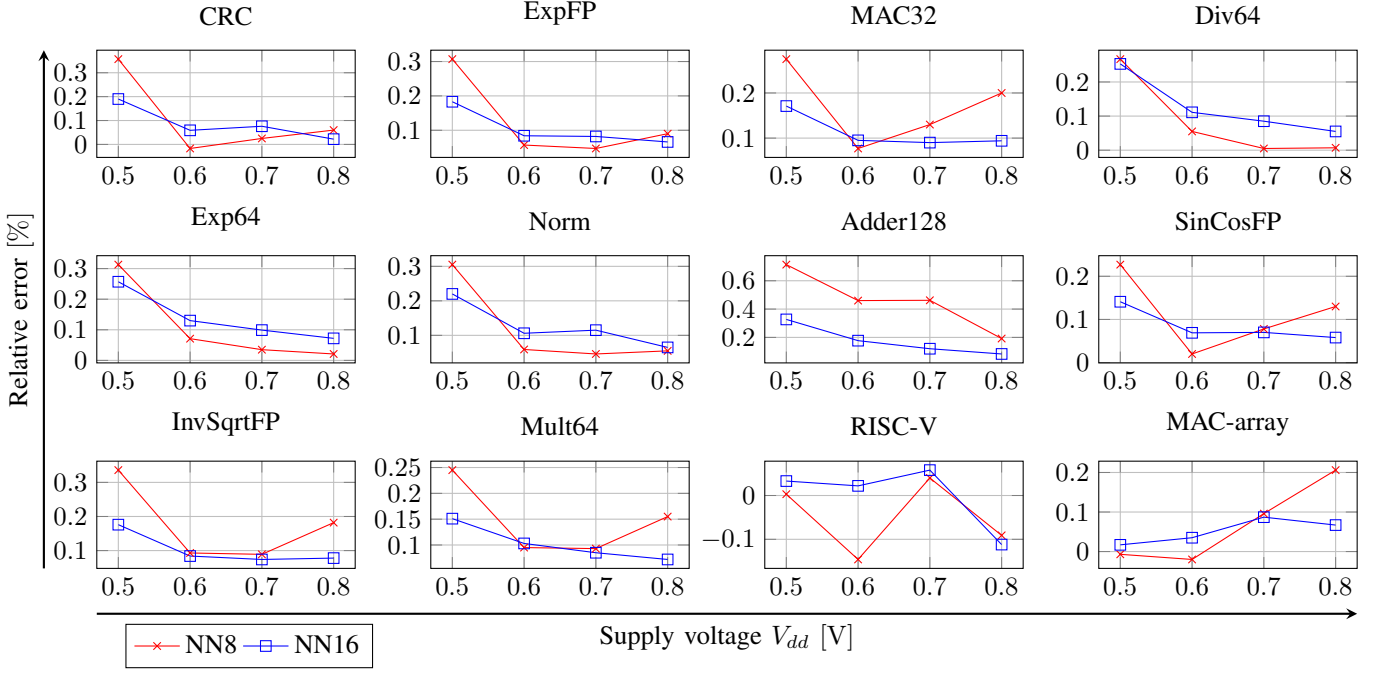

\section{Accuracy Evaluation at the Digital Level} \label{sec:digital}
In the following section we perform standard cell library characterization followed by digital circuit synthesis and static timing analysis. In addition to cell and circuit delays, we evaluate power and the speed of characterization, comparing NNs to the industry-standard BSIM-CMG.
\subsection{Standard Cell Library Characterization} \label{sec:expchar}
Standard cell library characterization is performed with commercial software using the setup described in Section \ref{sub:simsetup}. First, we investigate how NN-based transistor models perform with different supply voltages executing standard cell library characterization with default simulation setup, synthesis, and \sta for $V_{dd} \in [$\qty{0.5}{V}, \qty{0.6}{V}, \qty{0.7}{V}$]$, excluding low-precision setups. Fig. \ref{fig:sta_over_vdd} demonstrates that all models still maintain good accuracy in delay estimation. However, a noticeable trend is observed in a slight performance reduction for all NNs when the supply voltage is significantly lowered. This happens because capacitance behavior in the region near \qty{0.4}{V} was not learned as precisely as in the region near \qty{0.8}{V}, which was demonstrated previously in Fig. \ref{fig:cv_plots}. However, this error still remains negligible, especially for NN16, which experiences a significantly lower delay error increase than NN8 due to its larger size and model capacity, allowing it to cover a wider voltage region better. Additionally, NN16 demonstrates less chaotic behavior than NN8, since NN16 error follows a smooth and rather stable curve while NN8 fluctuates a lot.

\begin{figure*}
\centering
\begin{subfigure}[b]{0.3\textwidth} 
    \caption{NN16} 
    \includegraphics[height=3.6cm, keepaspectratio]{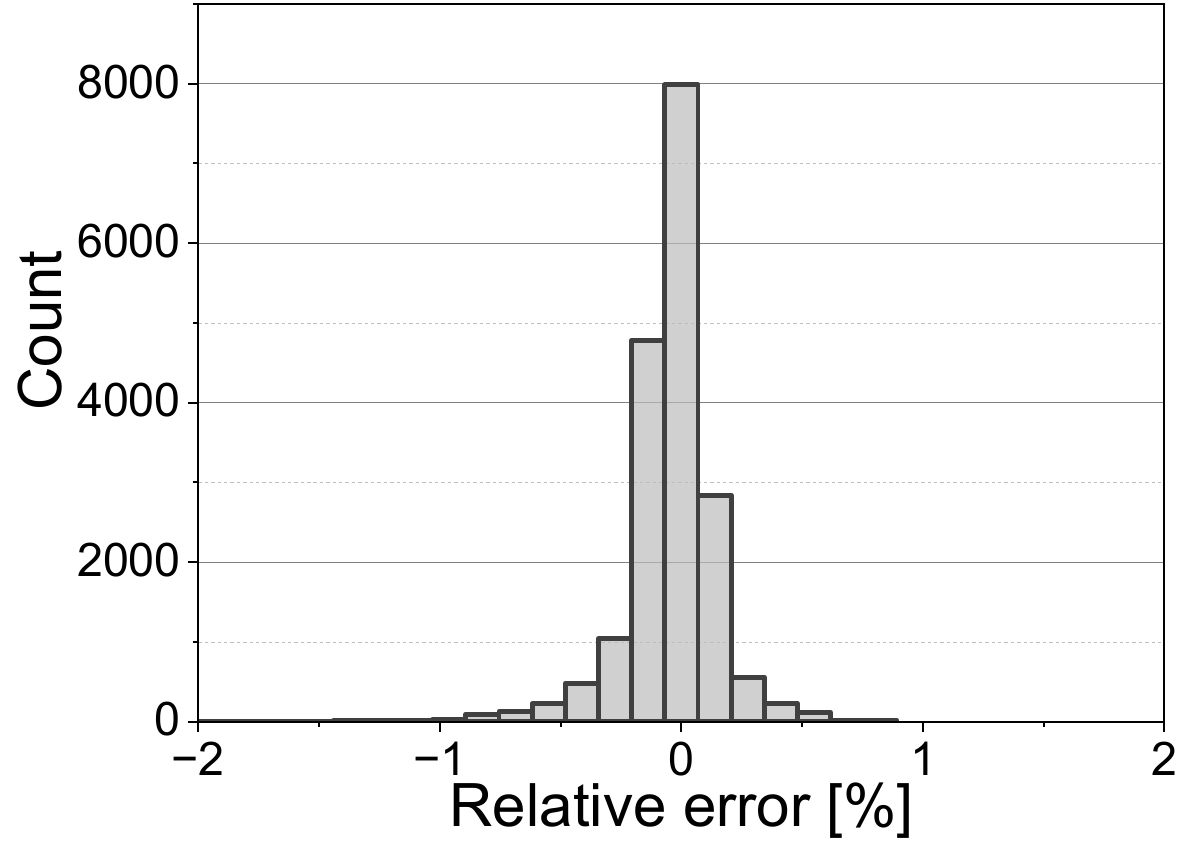}
    \label{fig:char_dist2}
\end{subfigure}\hspace{1.61em}%
\begin{subfigure}[b]{0.3\textwidth} 
    \caption{NN8} 
    \includegraphics[height=3.6cm, keepaspectratio]{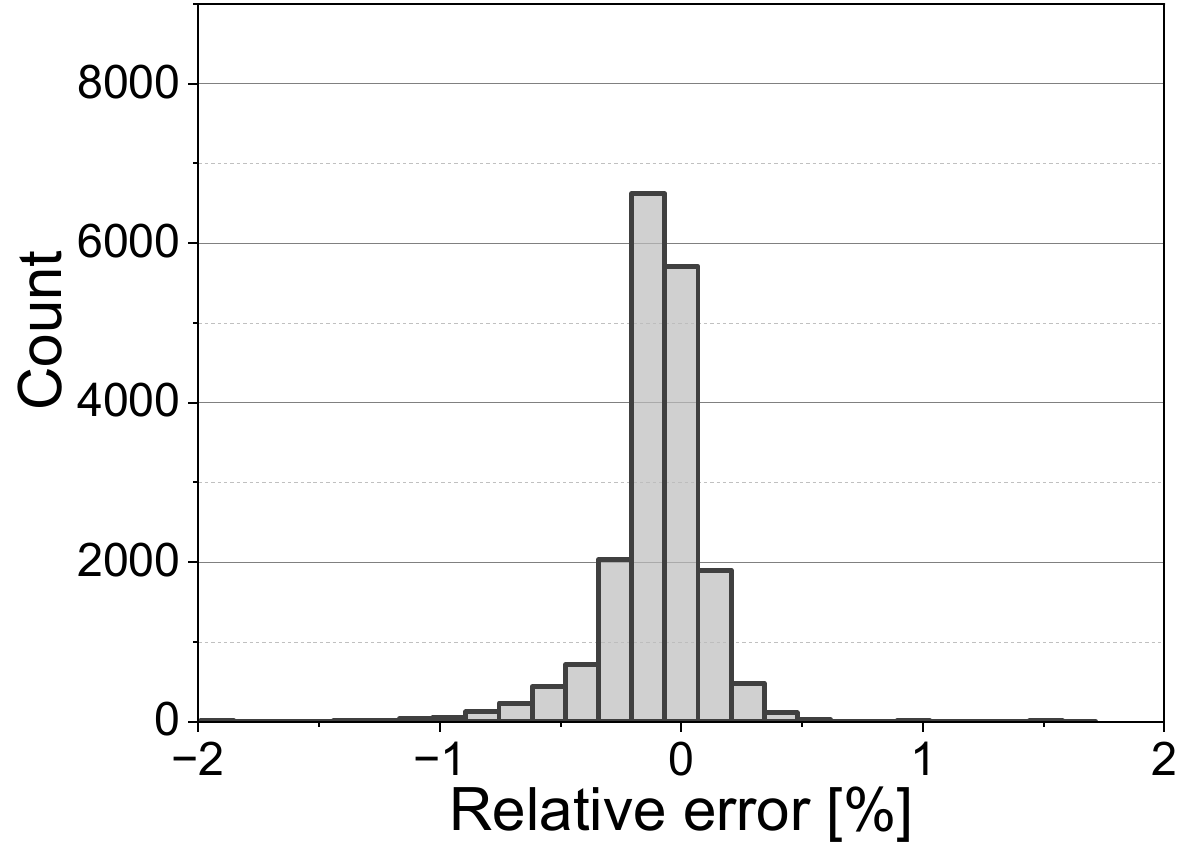}
    \label{fig:char_dist3}
\end{subfigure}\hspace{1.31em}%
\begin{subfigure}[b]{0.3\textwidth} 
    \caption{NN8 (LPS)} 
    \includegraphics[height=3.6cm, keepaspectratio]{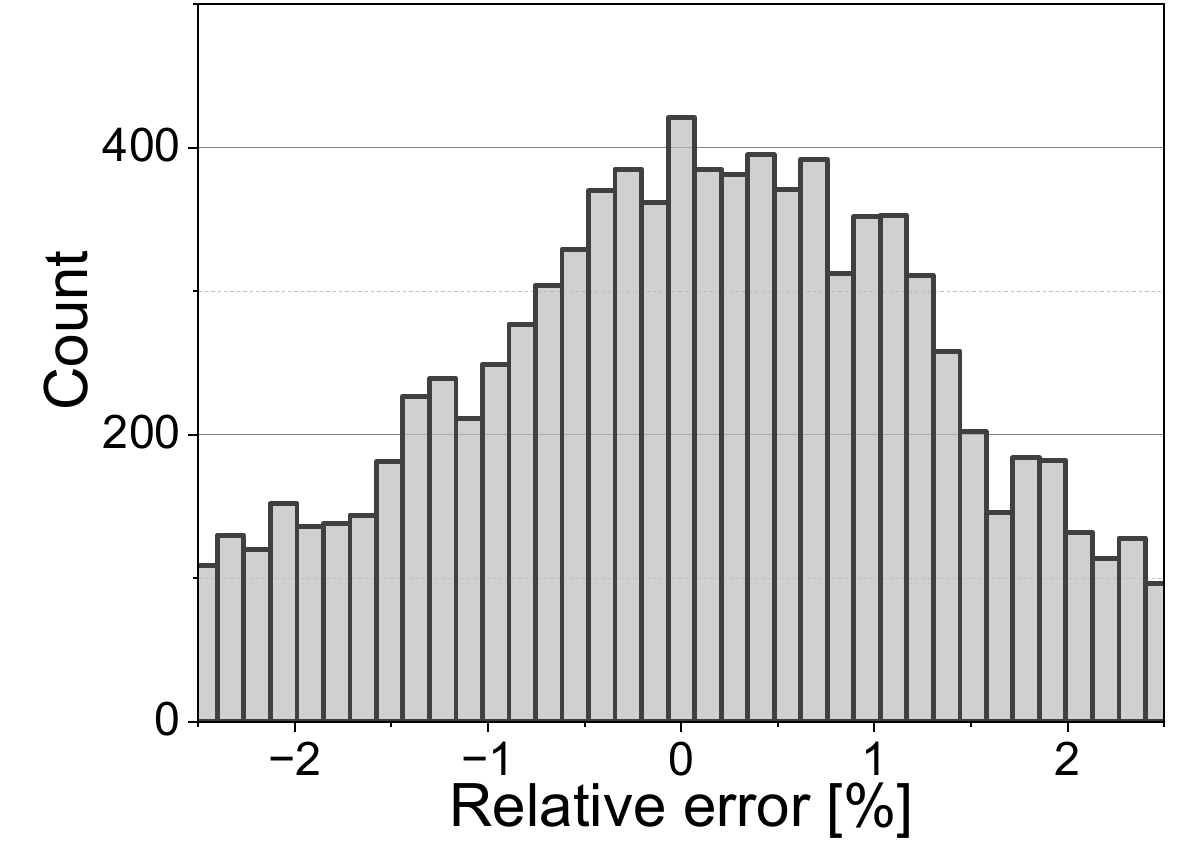}
    \label{fig:char_dist4}
\end{subfigure}

\medskip
    \caption{Distribution of relative errors for standard cell delays measured with respect to the BSIM-CMG. $V_{dd}$ = \qty{0.8}{V}. NN16 model demonstrates marginally better results over NN8 model. Standard cell library characterization performed with low-precision simulations shows significant deviations from the reference.} 
\label{fig:char_dist}
\end{figure*}

Since both NNs demonstrated remarkable delay prediction accuracy for the 16-bit divider circuit and $V_{dd}$ = \qty{0.7}{V}, the supply voltage for detailed analysis during characterization is set to \qty{0.8}{V}. Each cell type has several variations and incorporates multiple paths for delay measurement. All delay data from 1701 simulations is combined into a database (library). Fig. \ref{fig:char_dist} illustrates relative error distribution for 18620 delay entries (190 paths with 49 slew/capacitance combinations for rise and fall) with $0\%$ representing a perfect match with BSIM-CMG up to \qty{0.1}{ps}. NN16, shown in Fig. \ref{fig:char_dist2}, demonstrates the error distribution with almost 8000 perfect matches and an average relative error of $0.06\%$, while NN8 in Fig. \ref{fig:char_dist3} has fewer perfect matches with $0.08\%$ average error, which is still relatively low. Errors for the low-precision simulation setup, shown in Fig. \ref{fig:char_dist4}, fluctuate a lot and are relatively high compared to the reference. Multiple errors for NN8 (LPS) are also not shown, as for some x10 and x12 circuits, they exceeded even 100\%, making them unsuitable for any reasonable plot scale. The comparison of two simulation setups for NN8 indicates that the low-precision simulation is the main source of errors and should be employed with caution for standard cell library characterization, although it is perfectly capable of producing a good visual match for a circuit output, which was demonstrated earlier in Section \ref{sec:circuit}. Nevertheless, we will further investigate the performance of NN8 (LPS) in static timing analysis to determine how errors in standard cell delays are propagated into circuit delays.

\begin{table*}
\centering
\caption{Static timing analysis. Relative error (RE) is calculated with respect to BSIM-CMG. $V_{dd}$ = \qty{0.8}{V}.}
\label{table:sta}
\resizebox{0.85\textwidth}{!}{%
\begin{tabular}{c|c|cc|cc|cc}
\toprule
\multirow{2}{*}{Circuit} & BSIM-CMG & \multicolumn{2}{c|}{NN16} & \multicolumn{2}{c|}{NN8} & \multicolumn{2}{c}{NN8 (LPS)}\\
& Delay $[ns]$ & Delay $[ns]$ & RE [\%] & Delay $[ns]$ & RE [\%] & Delay $[ns]$ & RE [\%] \\ \midrule
CRC & \textcolor{black}{0.343551} & \textcolor{black}{0.343626} & \textcolor{black}{+0.022} & \textcolor{black}{0.343756} & \textcolor{black}{+0.060} & \textcolor{black}{0.341851} & \textcolor{black}{-0.495} \\
ExpFP & \textcolor{black}{3.177209} & \textcolor{black}{3.179308} & \textcolor{black}{+0.066} & \textcolor{black}{3.180068} & \textcolor{black}{+0.090} & \textcolor{black}{3.172469} & \textcolor{black}{-0.149} \\
MAC32 & \textcolor{black}{2.179060} & \textcolor{black}{2.181101} & \textcolor{black}{+0.094} & \textcolor{black}{2.183413} & \textcolor{black}{+0.200} & \textcolor{black}{2.176965} & \textcolor{black}{-0.096} \\
Div64 & \textcolor{black}{8.563543} & \textcolor{black}{8.568236} & \textcolor{black}{+0.055} & \textcolor{black}{8.564155} & \textcolor{black}{+0.007} & \textcolor{black}{8.540630} & \textcolor{black}{-0.268} \\
Exp64 & \textcolor{black}{8.669079} & \textcolor{black}{8.675332} & \textcolor{black}{+0.072} & \textcolor{black}{8.670883} & \textcolor{black}{+0.021} & \textcolor{black}{8.739084} & \textcolor{black}{+0.808} \\
Norm & \textcolor{black}{0.437502} & \textcolor{black}{0.437785} & \textcolor{black}{+0.065} & \textcolor{black}{0.437742} & \textcolor{black}{+0.055} & \textcolor{black}{0.438675} & \textcolor{black}{+0.268} \\
Adder128 & \textcolor{black}{3.355594} & \textcolor{black}{3.358387} & \textcolor{black}{+0.083} & \textcolor{black}{3.362043} & \textcolor{black}{+0.192} & \textcolor{black}{3.386376} & \textcolor{black}{+0.917} \\
SinCosFP & \textcolor{black}{4.819123} & \textcolor{black}{4.821896} & \textcolor{black}{+0.058} & \textcolor{black}{4.825410} & \textcolor{black}{+0.130} & \textcolor{black}{4.808432} & \textcolor{black}{-0.222} \\
InvSqrtFP & \textcolor{black}{4.086938} & \textcolor{black}{4.090143} & \textcolor{black}{+0.078} & \textcolor{black}{4.094364} & \textcolor{black}{+0.182} & \textcolor{black}{4.067212} & \textcolor{black}{-0.483} \\
Mult64 & \textcolor{black}{4.752072} & \textcolor{black}{4.755509} & \textcolor{black}{+0.072} & \textcolor{black}{4.759443} & \textcolor{black}{+0.155} & \textcolor{black}{4.754639} & \textcolor{black}{+0.054} \\
RISC-V & \textcolor{black}{1.482343} & \textcolor{black}{1.480690} & \textcolor{black}{-0.112} & \textcolor{black}{1.480992} & \textcolor{black}{-0.091} & \textcolor{black}{1.486655} & \textcolor{black}{+0.292} \\
MAC-array & \textcolor{black}{0.949363} & \textcolor{black}{0.949995} & \textcolor{black}{+0.067} & \textcolor{black}{0.951319} & \textcolor{black}{+0.206} & \textcolor{black}{0.951861} & \textcolor{black}{+0.263} \\
\bottomrule
\end{tabular}}
\end{table*}

\subsection{Standard Cell Library Characterization Speedup} \label{sec:simspeed}
\begin{figure}[t]
    \centering
    \begin{tikzpicture}
        \begin{axis}[
            ybar,
            bar width=25pt,
            width=0.98\columnwidth,
            height=4cm,
            ylabel={Time [min]},
            symbolic x coords={BSIM-CMG, NN16, NN8, NN8 (LPS)},
            xtick=data,
            ymin=0, ymax=70,
            enlarge x limits=0.4,
            nodes near coords,
            every node near coord/.append style={
                font=\small\color{black},
                yshift=2pt
            },
            ymajorgrids=true,
            grid style=dashed,
            xticklabel style={font=\footnotesize},
            axis lines*=left,
        ]
            \addplot[fill=blue!30, draw=blue!80] coordinates {
                (BSIM-CMG, 55) 
                (NN16, 7) 
                (NN8, 4.1)
                (NN8 (LPS), 3.2)
            };

            \node[anchor=south, font=\footnotesize\bfseries, color=red!70!black] at (axis cs:NN16, 18) {$\sim 8\times$};
            \node[anchor=south, font=\footnotesize\bfseries, color=red!70!black] at (axis cs:NN8, 18) {$\sim 13\times$};
            \node[anchor=south, font=\footnotesize\bfseries, color=red!70!black] at (axis cs:{NN8 (LPS)}, 18) {$\sim 17\times$};
            \node[anchor=south, font=\footnotesize\bfseries, color=red!70!black] at ([xshift=0pt]axis cs:NN8, 27) {Speedup};
            
        \end{axis}
    \end{tikzpicture}
    \caption{NN-based transistor models show a significant reduction in time required for standard cell library characterization, with NN16 being approximately $8\times$ faster than BSIM-CMG.}
    \label{fig:char_time_chart}
\end{figure}
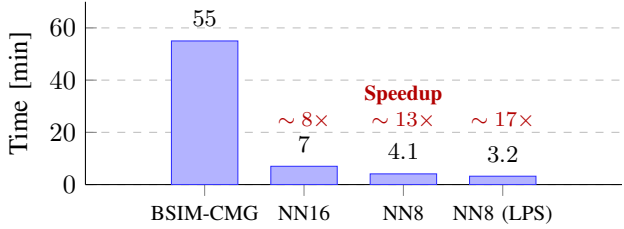

The runtime for the standard cell library characterization, as shown in Fig. \ref{fig:char_time_chart}, follows a similar pattern, which was observed for circuits in Section \ref{sec:circuit_speed}. NN8 completes all simulations in the shortest amount of time of 4.1 minutes, while Verilog-A BSIM-CMG takes 55 minutes, which is the longest for the same task. Since computational throughput plays a minor role in the standard cell library characterization, NN16 is relatively close to NN8, with a runtime of 7 minutes. In contrast to the 16-bit divider circuit simulation, where the number of calculations is the major factor, characterization speed greatly depends on the amount of code that needs to be compiled. For a single large circuit, Verilog-A code is compiled only once, and then a significant number of calculations is executed. The complexity of standard cells is much lower than that of the circuits used in Section \ref{sec:circuit}. Consequently, a simulation of a single cell will be performed almost instantly and will not be significantly affected by a low-precision simulation setup. However, all Verilog-A implementations require compiling the transistor model for each netlist, meaning that standard cell library characterization runtime greatly depends on the size of the code. With the Verilog-A BSIM-CMG implementation having the largest code base, compiling it multiple times during characterization takes a considerable amount of time. Standard cell library characterization speed with NN-based models is significantly lower than that of the Verilog-A BSIM-CMG due to the simplistic nature of the NN code, which primarily contains matrix multiplications. Notably, compilation time for the Verilog-A code is not affected by the low-precision simulation setup and remains constant for the same implementation, keeping the benefit really low compared to analog level circuit simulations in Section \ref{sec:circuit_speed}.

\subsection{Static Timing Analysis} \label{sec:expsta}
Static timing analysis (STA) is performed with a commercial software using libraries (databases) generated during the standard cell library characterization described in Section \ref{sec:expchar}. Circuit synthesis is done once and is based on the BSIM-CMG library. Then, all 12 synthesized circuits are used to calculate the respective delays for each transistor model. Table \ref{table:sta} shows that NN16, on average, outperforms NN8 and produces delay measurements closer to the reference ones from BSIM-CMG. Additionally, the NN16 model demonstrates rather stable results, with the relative error ranging from $-0.112\%$ to $0.094\%$, while the NN8 model fluctuates between $-0.091\%$ and $0.206\%$. Nevertheless, NN8, despite having much higher errors than NN16 for various transistor characteristics (see Table \ref{table:accuracy}), still shows remarkable delay prediction accuracy. Given the high precision across all 12 circuits, the results indicate that all NN-based transistor models perform with an accuracy level very close to the reference BSIM-CMG. In contrast to other setups, NN8 (LPS) exhibits significant instability and higher delay errors, reaching $0.917\%$, which is primarily caused by the low simulation precision itself.

\begin{table*}
\centering
\caption{Circuit power analysis with NN-based transistor model. Absolute values are given for NN16, and relative errors (RE) are calculated with respect to BSIM-CMG.}
\label{table:power_analysis}
\resizebox{0.9\textwidth}{!}{%
\begin{tabular}{l|cc|cc|cc|cc}
\toprule
\multirow{2}{*}{Circuit} & \multicolumn{2}{c|}{Net switching power} & \multicolumn{2}{c|}{Cell internal power} & \multicolumn{2}{c|}{Cell leakage power} & \multicolumn{2}{c}{Total power} \\
 & Value [W] & RE [\%] & Value [W] & RE [\%] & Value [W] & RE [\%] & Value [W] & RE [\%] \\ \midrule
CRC & \textcolor{black}{7.032e-05} & \textcolor{black}{-0.043} & \textcolor{black}{4.395e-05} & \textcolor{black}{+0.023} & \textcolor{black}{1.251e-07} & \textcolor{black}{+0.160} & \textcolor{black}{1.144e-04} & \textcolor{black}{+0.000} \\
ExpFP & \textcolor{black}{1.664e-04} & \textcolor{black}{+0.000} & \textcolor{black}{1.570e-04} & \textcolor{black}{+0.512} & \textcolor{black}{3.811e-06} & \textcolor{black}{+0.184} & \textcolor{black}{3.271e-04} & \textcolor{black}{+0.214} \\
MAC32 & \textcolor{black}{5.252e-04} & \textcolor{black}{+0.019} & \textcolor{black}{7.313e-04} & \textcolor{black}{+0.647} & \textcolor{black}{9.299e-06} & \textcolor{black}{+0.216} & \textcolor{black}{1.266e-03} & \textcolor{black}{+0.397} \\
Div64 & \textcolor{black}{6.088e-04} & \textcolor{black}{-0.033} & \textcolor{black}{4.116e-04} & \textcolor{black}{+0.195} & \textcolor{black}{3.661e-05} & \textcolor{black}{+0.192} & \textcolor{black}{1.057e-03} & \textcolor{black}{+0.095} \\
Exp64 & \textcolor{black}{2.079e-03} & \textcolor{black}{+0.000} & \textcolor{black}{1.406e-03} & \textcolor{black}{+0.214} & \textcolor{black}{3.653e-05} & \textcolor{black}{+0.192} & \textcolor{black}{3.521e-03} & \textcolor{black}{+0.057} \\
Norm & \textcolor{black}{1.539e-05} & \textcolor{black}{-0.065} & \textcolor{black}{7.716e-06} & \textcolor{black}{+0.065} & \textcolor{black}{4.667e-08} & \textcolor{black}{+0.215} & \textcolor{black}{2.316e-05} & \textcolor{black}{+0.000} \\
Adder128 & \textcolor{black}{2.500e-05} & \textcolor{black}{-0.080} & \textcolor{black}{1.959e-05} & \textcolor{black}{-0.659} & \textcolor{black}{6.283e-07} & \textcolor{black}{+0.207} & \textcolor{black}{4.522e-05} & \textcolor{black}{-0.309} \\
SinCosFP & \textcolor{black}{4.627e-05} & \textcolor{black}{+0.000} & \textcolor{black}{4.602e-05} & \textcolor{black}{+0.590} & \textcolor{black}{5.200e-06} & \textcolor{black}{+0.212} & \textcolor{black}{9.749e-05} & \textcolor{black}{+0.298} \\
InvSqrtFP & \textcolor{black}{6.698e-05} & \textcolor{black}{-0.030} & \textcolor{black}{6.722e-05} & \textcolor{black}{+0.164} & \textcolor{black}{1.221e-06} & \textcolor{black}{+0.246} & \textcolor{black}{1.354e-04} & \textcolor{black}{+0.074} \\
Mult64 & \textcolor{black}{8.945e-04} & \textcolor{black}{-0.011} & \textcolor{black}{1.015e-03} & \textcolor{black}{+0.396} & \textcolor{black}{2.074e-05} & \textcolor{black}{+0.193} & \textcolor{black}{1.930e-03} & \textcolor{black}{+0.208} \\
RISC-V & \textcolor{black}{9.750e-05} & \textcolor{black}{+0.062} & \textcolor{black}{7.231e-04} & \textcolor{black}{+0.865} & \textcolor{black}{6.665e-06} & \textcolor{black}{+0.226} & \textcolor{black}{8.272e-04} & \textcolor{black}{+0.755} \\
MAC-array & \textcolor{black}{4.010e-03} & \textcolor{black}{+0.025} & \textcolor{black}{1.650e-02} & \textcolor{black}{+1.227} & \textcolor{black}{1.618e-04} & \textcolor{black}{+0.186} & \textcolor{black}{2.060e-02} & \textcolor{black}{+0.488} \\
\bottomrule

\end{tabular}}
\end{table*}

\begin{figure}
\centering
\begin{subfigure}[b]{0.24\textwidth} 
    \caption{NN16} 
    \includegraphics[height=3.1cm, keepaspectratio]{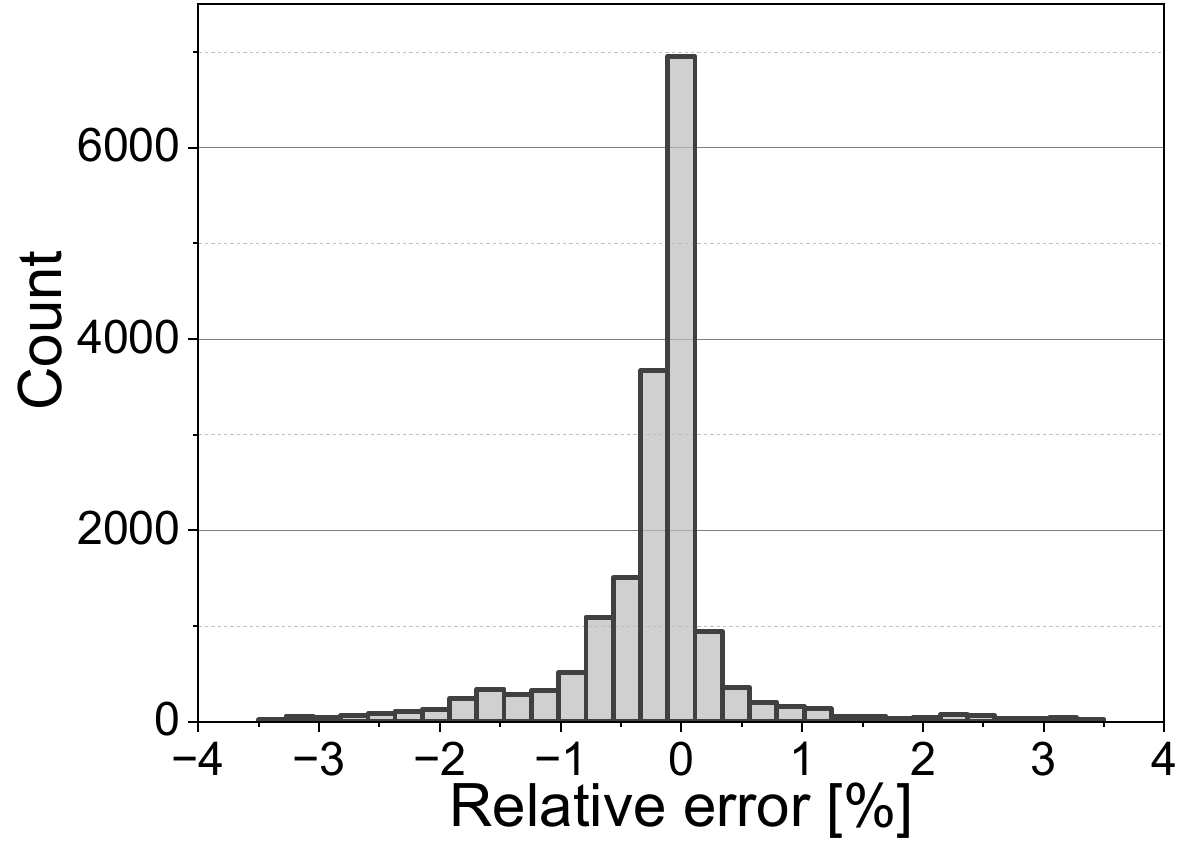}
    \label{fig:char_power2}
\end{subfigure}\hspace{0.01em}%
\begin{subfigure}[b]{0.24\textwidth} 
    \caption{NN8} 
    \includegraphics[height=3.1cm, keepaspectratio]{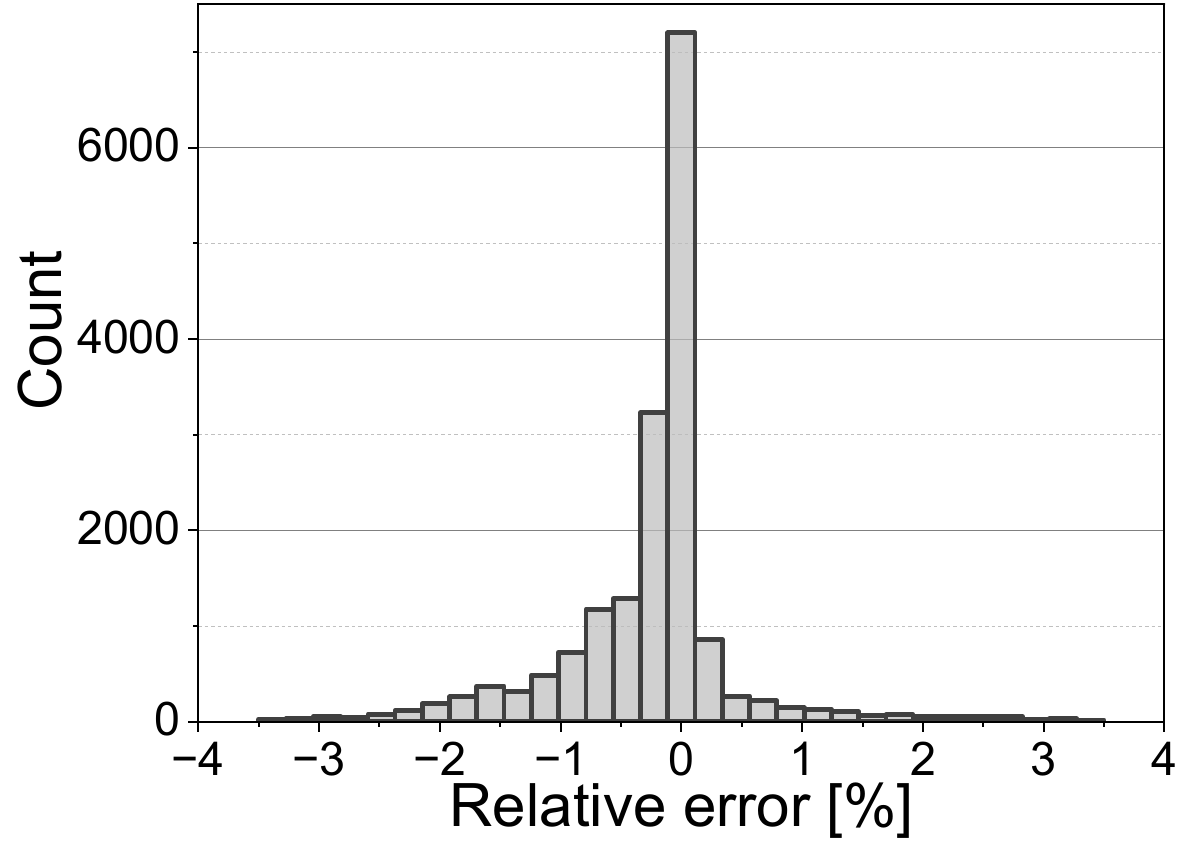}
    \label{fig:char_power3}
\end{subfigure}\hspace{0.31em}%

\medskip
    \caption{Distribution of relative errors for cell internal power measured with respect to the BSIM-CMG. $V_{dd}$ = \qty{0.8}{V}. NN8 error distribution looks visually similar to the NN16 one. However, the average error for NN16 is lower than that for NN8 due to higher accuracy.} 
\label{fig:char_power}
\end{figure}

\subsection{Circuit Power Signoff Analysis} \label{sec:power}
In addition to timing data, standard cell library characterization provides information on the internal cell power. Fig. \ref{fig:char_power} shows the relative error distribution for the same 18620 entries employed in timing analysis. Even though NN8 seems to have a very similar error distribution as NN16, the average error of $0.3275\%$ for NN16 is lower than $0.3324\%$, demonstrated by NN8. Since the cell power library is used to estimate the power of complex circuits, it is more beneficial to employ a model with stable and consistent behavior, which is capable of producing smooth transitions between look-up table (LUT) elements for accurate circuit power estimation. The effect of model stability can be easily noticed in the next experiment.

Circuit analysis provides multiple power characteristics: net switching power, cell internal power, and cell leakage power. Table \ref{table:power_analysis} illustrates the power report for a 16-neuron NN-based transistor model and compares its power estimation with BSIM-CMG. Net switching power estimation is the most accurate among the three power components and shows an ideal match for three circuits, with an error below $0.08\%$ for the other ones, indicating a good gate capacitance and ON-current match between two transistor models. Cell leakage power, being relatively low compared to other powers, does not contribute significantly to the total circuit power. Nevertheless, the NN-based transistor model was able to estimate it with high accuracy and relative errors below $0.25\%$ compared to the industry-standard BSIM-CMG.

In contrast to other power components, cell internal power demonstrates the largest relative error, ranging from $-0.659\%$ to $1.227\%$, since it is extremely sensitive to any numerical differences between models. Internal power is primarily influenced by CMOS short circuit conditions and slew propagation, which lead to highly nonlinear behavior, instability, and overall difficulty during estimation. Additionally, circuit power estimation is based on the LUTs from individual cells, which introduces another level of computational errors. All these factors result in error accumulation during different stages of calculation, leading to a more noticeable deviation in the cell internal power. Total power, as a sum of all three powers, also exhibits higher error than switching power on average, but this deviation is primarily driven by the high error in the cell internal power. Notably, power estimation performed with an 8-bit NN-based transistor model exhibits only slightly higher errors for net switching power and similar ones for cell leakage power, with the maximum errors of $0.27\%$ and $0.25\%$, respectively. However, the relative error for cell internal power, estimated with the NN8 model, ranges from $0.43\%$ to $4.8\%$ and is spread uniformly across this range for all 12 circuits, highlighting the instability of smaller NNs and their struggle with modeling highly nonlinear transistor behavior.

\section{Conclusion} \label{sec:conclusion}
In this paper, we demonstrated that NN-based transistor models can provide high-quality results for standard cell library characterization and further circuit analysis, while achieving significantly higher execution and convergence speeds compared to the industry-standard Verilog-A implementation. Even though circuit-level analysis involves multiple levels of simulations, NN-based transistor models were able to maintain remarkable performance when employed as a foundation in complex multi-stage tasks. Given the right architecture and size, NN-based transistor models open up the possibility of performing precise circuit simulations for any emerging technology without the need to describe underlying physical processes, significantly accelerating the research and electronic design automation flow, especially when employed with GPU acceleration.

\section*{Acknowledgement}
Authors would like to thank Anirban Kar, Shivendra Singh Parihar, and Girish Pahwa for insightful discussions regarding transistor compact modeling in Verilog-A; Mahdi Benkhelifa and Albi Mema for valuable support during standard cell library characterization and circuit analysis.



\bibliographystyle{IEEEtran}
\bibliography{references}

\begin{IEEEbiography}[{\includegraphics[width=1in,height=1.25in,clip,keepaspectratio]{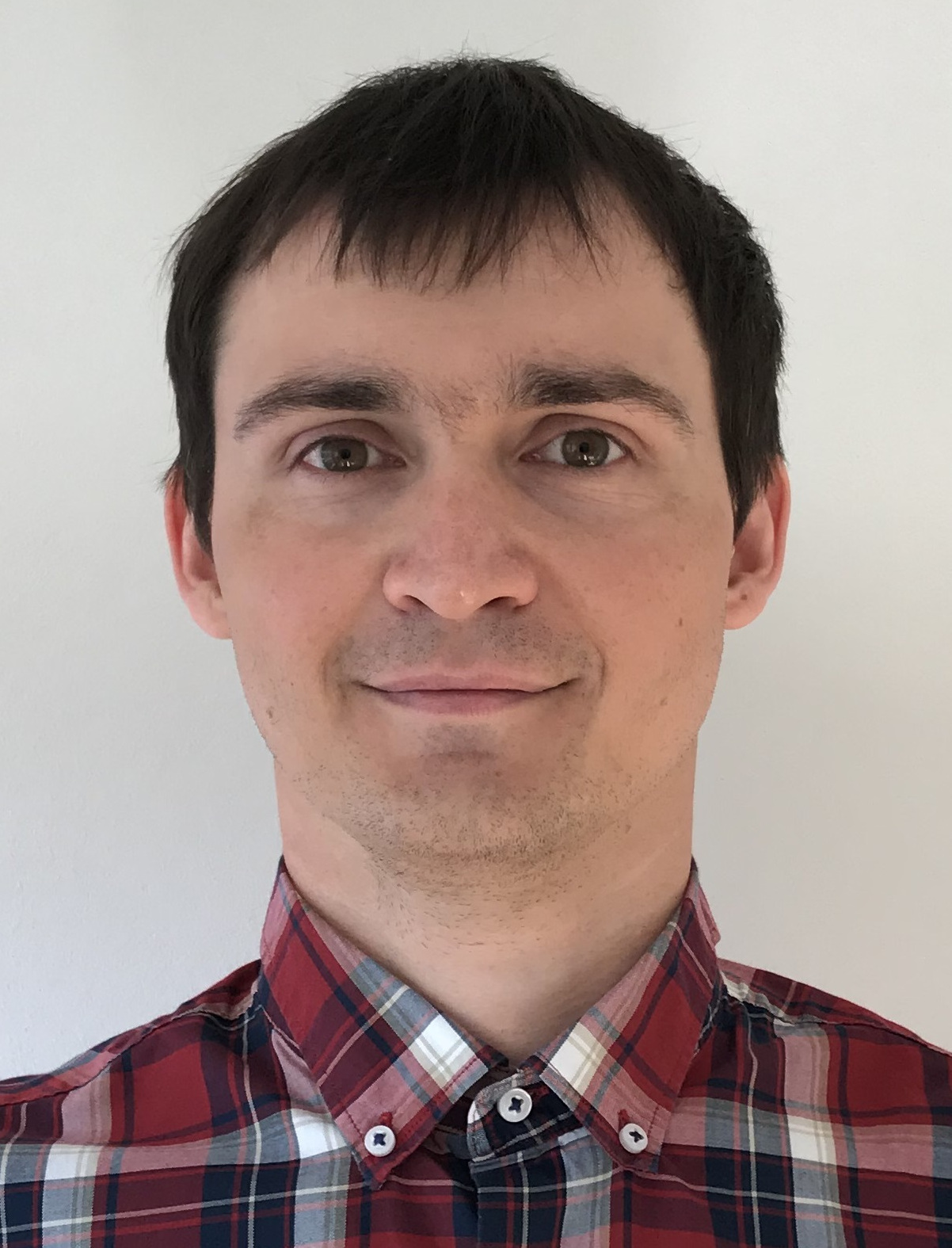}}] {Rodion Novkin} is a Doctoral Researcher at the Chair of AI Processor Design within the Technical University of Munich (TUM). He is also with the Munich Institute of Robotics and Machine Intelligence in Germany. He received the B.Sc. in Robotics and Mechatronics from the Bauman Moscow State Technical University, Russian Federation, in 2013 and the M.Sc. in Automation and Robotics from the Technical University Dortmund, Germany, in 2021. He is currently working towards the Ph.D. degree at the Chair of AI Processor Design within TUM. His research interests include low-power deep neural networks, machine learning for embedded systems and computer-aided design.
\end{IEEEbiography}
\begin{IEEEbiography}[{\includegraphics[width=1in,height=1.25in,clip,keepaspectratio]{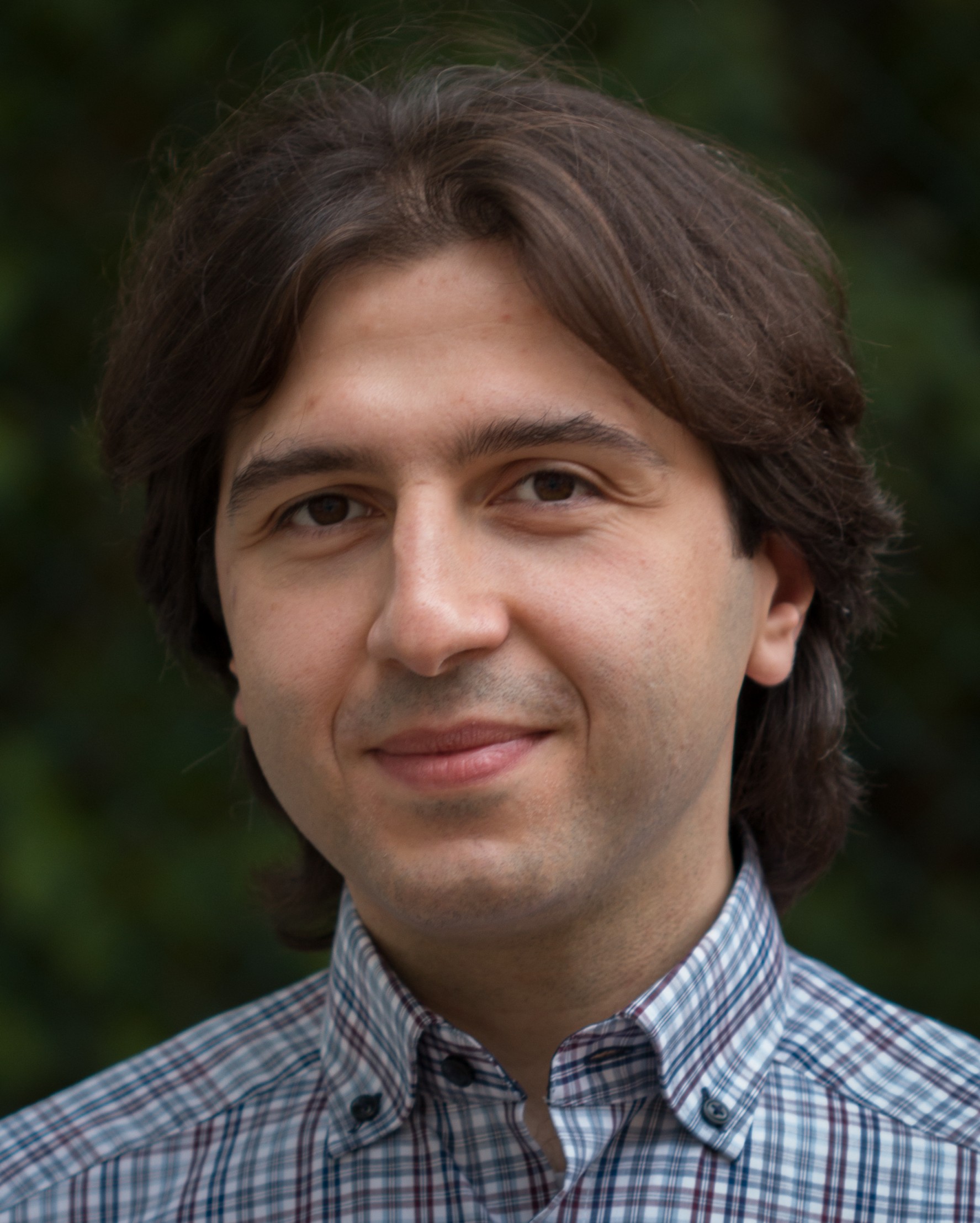}}] {Hussam Amrouch}(S'11-M'15) Professor heading the Chair of AI Processor Design within the Technical University of Munich. He is the Founding Director of the Munich Advanced-Technology Center for AI Chips.
He is the head of the Brain-inspired Computing within the Munich Institute of Robotics and Machine Intelligence. He is the head of the Semiconductor Test and Reliability at the Uni. of Stuttgart. He received his Ph.D. degree with the highest distinction (Summa cum laude) from KIT in 2015. He holds 10x HiPEAC Paper Awards. He has more than $320$ publications in multidisciplinary research areas across the entire computing stack, starting from semiconductor physics to circuit design and EDA.
\end{IEEEbiography}
\end{document}

%% file: glossar.tex
\usepackage{xspace}
\xspaceaddexceptions{)}
\usepackage[acronyms]{glossaries}
\hypersetup{hidelinks}

\newglossaryentry{vdd}{
	name={\ensuremath{V_{\text{DD}}}},
	description={supply voltage},
	symbol=Vdd,
	sort=vdd}
\newglossaryentry{vth}{
	name={\ensuremath{V_{\text{TH}}}},
	description={threshold voltage},
	symbol=Vth,
	sort=vth}
\newglossaryentry{gnd}{
	name={GND},
	description={El. ground},
	symbol=GND}

\newacronym{aef}{AEF}{average electric field}
\newacronym{dnn}{DNN}{deep neural network}
\newacronym{nn}{NN}{Neural network}
\newacronym{kan}{KAN}{Kolmogorov-Arnold network}
\newacronym{fkan}{FKAN}{Fourier KAN}
\newacronym{kart}{KART}{Kolmogorov-Arnold representation theorem}
\newacronym{cpn}{CPN}{current/charge prediction network}
\newacronym{pgn}{PGN}{parameter generator network}
\newacronym{gnn}{GNN}{graph neural network}
\newacronym{rnn}{RNN}{reccurrent neural network}
\newacronym{cnn}{CNN}{convolutional neural network}
\newacronym{aspp}{ASPP}{atrous spatial pyramid pooling}
\newacronym{mape}{MAPE}{mean absolute percentage error}
\newacronym{rmse}{RMSE}{root mean square error}
\newacronym{dtco}{DTCO}{design technology co-optimization}
\newacronym{sta}{STA}{static timing analysis}
\newacronym{iou}{IoU}{intersection over union}
\newacronym{fom}{FoM}{figure of merit}
\newacronym{miou}{MIoU}{mean intersection over union}
\newacronym{espp}{EsPP}{electrostatic potential profile}
\newacronym{mse}{MSE}{mean square error}
\newacronym{bjt}{BJT}{bipolar junction transistor}
\newacronym{ml}{ML}{Machine learning}
\newacronym{mlp}{MLP}{multi-layer perceptron}

\newacronym{rnet}{R-Net}{resistor network}

\newacronym{tcad}{TCAD}{Technology CAD}
\newacronym{cad}{CAD}{computer-aided design}
\newacronym{spice}{SPICE}{Simulation Program with Integrated Circuit Emphasis}
\newacronym{hspice}{HSPICE}{PrimeSim HSPICE}
\newacronym{ddic}{DDIC}{display driver integration circuit}
\newacronym{ss}{SS}{sub-threshold slope}

\newacronym{fdsoi}{FDSOI}{Fully-Depleted Silicon on Insulator}
\newacronym{fefet}{FeFET}{Ferroelectric FET}
\newacronym{finfet}{FinFET}{fin field-effect transistor}
\newacronym{fe}{FE}{ferroelectric}
\newacronym{mlc}{MLC}{multi-level cell}

\newacronym{nls}{NLS}{nucleation-limited switching}
\newacronym{ds}{DS}{Domain Stochasticity}

\newacronym{dtd}{DTD}{Device-to-Device}
\newacronym{ctc}{CTC}{Cycle-to-Cycle}
\newacronym{mac}{MAC}{multiply-accumulate}

\newacronym{stt}{STT}{Spin Transfer Torque}
\newacronym{rram}{ReRAM}{Resistive RAM}
\newacronym{pde}{PDE}{partial differential equation}

\newcommand{\nn}{\gls{nn}\xspace}

\newcommand{\gnn}{\gls{gnn}\xspace}

\newcommand{\kan}{\gls{kan}\xspace}

\newcommand{\mape}{\gls{mape}\xspace}
\newcommand{\rmse}{\gls{rmse}\xspace}

\newcommand{\sta}{\gls{sta}\xspace}

\newcommand{\mlp}{\gls{mlp}\xspace}

\newcommand{\vds}{\ensuremath{V_{\text{D}}}\xspace}
\newcommand{\vgs}{\ensuremath{V_{\text{G}}}\xspace}
\newcommand{\qd}{\ensuremath{Q_{\text{D}}}\xspace}
\newcommand{\qs}{\ensuremath{Q_{\text{S}}}\xspace}
\newcommand{\qg}{\ensuremath{Q_{\text{G}}}\xspace}
\newcommand{\qb}{\ensuremath{Q_{\text{B}}}\xspace}
\newcommand{\gm}{\ensuremath{g_{\text{m}}}\xspace}
\newcommand{\gds}{\ensuremath{g_{\text{DS}}}\xspace}

\newcommand{\ids}{\ensuremath{I_{\text{D}}}\xspace}